# $Nb_3Sn$ coating of SRF cavity by co-sputtering from a composite target

Md Sharifuzzaman Shakel[1,2], Grigory V. Eremeev[3], Anne-Marie Valente-Feliciano[4], Uttar Pudasaini[4], and Hani E. Elsayed-Ali[1,2*]

[1]*Department of Electrical and Computer Engineering, Old Dominion University, Norfolk, VA 23529, USA*
[2]*Applied Research Center, 12050 Jefferson Avenue, Newport News, VA 23606, USA*
[3]*Fermi National Accelerator Laboratory, Batavia, IL 60510, USA*
[4]*Thomas Jefferson National Accelerator Facility, Newport News, VA 23606, USA*

We deposited $Nb_3Sn$ film on the inner surface of a 2.6 GHz Nb superconducting radiofrequency (SRF) cavity by co-sputtering using a composite of Nb and Sn tube targets in a DC cylindrical magnetron sputtering system, followed by thermal annealing of the coated cavity. An aluminum mockup cavity, replicating a 2.6 GHz Nb SRF cavity geometry, was utilized to optimize deposition parameters, resulting in co-sputtered Nb-Sn films with Sn content of 32 – 42 at. % on the beam tubes and equator positions. Several annealing conditions were investigated to improve the surface homogeneity of the $Nb_3Sn$ film. The best co-sputtered Nb-Sn film was achieved after annealing at 600 °C for 6 h, followed by annealing at 950 °C for 1 h. The best process was applied to a Nb cavity, which was RF tested in a cryogenic dewar. RF testing of the $Nb_3Sn$-coated cavity demonstrated a superconducting transition temperature ($T_c$) close to the highest reported, at 17.78 K. The $Nb_3Sn$ cavity underwent a light Sn recoating process, followed by additional RF testing, resulting in enhancement of the RF performance, primarily due to improved surface homogeneity of the $Nb_3Sn$ coating.

* Corresponding author: helsayed@odu.edu

## 1. Introduction

Due to its highest superconducting critical temperature ($T_c$) and highest superheating field ($H_{sh}$) among pure metals, along with favorable mechanical properties for the fabrication of complex geometries, niobium (Nb) has long been the state-of-the-art material for superconducting radio-frequency (SRF) accelerator cavities [1,2]. The performance of bulk Nb cavities has improved significantly by addressing limiting factors such as multipacting, field emission, thermal quench, and high-field Q-slope [2]. While ongoing efforts to further enhance cavity performance through various techniques such as electropolishing, nitrogen doping and infusion, oxygen-based treatments, and medium temperature baking have shown promise, Nb cavities are approaching their intrinsic performance limits in terms of surface resistance and maximum accelerating gradient [3–10]. Therefore, there has been a significant interest in alternative materials with enhanced superconducting properties for developing advanced SRF technologies [11–13]. Since the top few hundred nanometers of the cavity surface are crucial to its RF performance, depositing a thin film of material possessing superconducting properties superior to Nb on cavity structures is considered the next step for advanced SRF accelerator cavities [14].

$Nb_3Sn$, an A15-type superconductor, is a promising alternative to Nb due to its significantly higher $T_c$ of 18.3 K and $H_{sh}$ of about 400 mT, both nearly double that of Nb [15]. Bulk $Nb_3Sn$ is brittle; however, due to the closely matched thermal expansion coefficients of $Nb_3Sn$ and Nb, $Nb_3Sn$ is suitable as a thin film coating for Nb SRF cavities. $Nb_3Sn$-coated SRF cavities operating at 4.2 K can attain quality factors ($Q_0$) comparable to conventional Nb cavities at 2.0 K, offering significant reductions in cryogenic system complexity and operating costs [16]. Multiple research groups, including those at Fermilab, Jefferson Lab, Cornell University, and others, have been developing techniques for coating $Nb_3Sn$ into SRF cavities [17–20]. Sn vapor diffusion method has been the most successful approach for $Nb_3Sn$ coating of Nb SRF cavities and has demonstrated the highest effectiveness in terms of the RF performance of the cavity. $Nb_3Sn$ fabrication using vapor diffusion involves a two-step process: a nucleation step around about 500 °C, where $SnCl_2$ forms Sn sites on the Nb surface, followed by a coating step at 1100 – 1200 °C, during which Sn vapor reacts with the Nb substrate to form $Nb_3Sn$ [17]. Fermilab reported that a $Nb_3Sn$-coated single-cell 1.3 GHz Nb cavity developed using the Sn vapor diffusion process has demonstrated a $Q_0$ of about $10^{10}$ at 4.4 K, reaching an accelerating gradient ($E_{acc}$) up to 24 MV/m [17]. However, $Nb_3Sn$ surfaces produced via Sn vapor diffusion can possess several imperfections. For example, patchy regions with significantly reduced thickness compared to the surrounding $Nb_3Sn$ coating are observed [21]. These regions can cause increased surface resistance, leading to early magnetic flux penetration and quenching of the cavity at lower fields [22]. Additionally, Sn-deficient areas on the coating were observed, resulting in a suppression of the $T_c$ of the $Nb_3Sn$ layer [23]. These features develop during $Nb_3Sn$ vapor diffusion coating and are under investigation to be understood and eliminated.

One alternative method for $Nb_3Sn$ film fabrication is magnetron sputtering, which offers improved control over deposition rates and stoichiometry and reduced substrate temperature during film growth. Magnetron sputtering has been employed in several configurations to fabricate $Nb_3Sn$ on flat substrates, including sequential sputtering of Nb and Sn layers followed by annealing, co-sputtering of Nb and Sn, and sputtering from a single stoichiometric target [24–26]. In the sequential layer sputtering approach, precise control of the individual layer thickness and the thickness ratio of Nb and Sn layers is required to achieve the desired film composition [27]. However, the co-sputtering approach allows simultaneous sputtering of Nb and Sn, enabling more direct control of the film composition by adjusting the sputtered atomic fluxes of the individual elements. Co-sputtering can also allow incorporation of additional elements into $Nb_3Sn$ films, which can serve as artificial pinning centers to enhance superconducting properties by restricting fluxon motion [28,29]. Schäfer et al. demonstrated the fabrication of crystalline $Nb_3Sn$ by co-sputtering on flat Cu substrates at a low substrate temperature of 435 $^0$C, achieving a $T_c$ of 16.35 K [30]. Sayeed et al. demonstrated co-sputtering deposition of Nb and Sn on flat Nb substrates at room-temperature, followed by annealing at 950 $^0$C, resulting in crystalline $Nb_3Sn$ films with large grains and $T_c$ of 17.61 K [25]. However, extending the co-sputtering approach to SRF cavities introduces additional challenges associated with complex cavity geometry, including spatially varying plasma-surface interactions and nonuniform transport of sputtered atoms towards different regions of the cavity.

We present the results of Nb-Sn co-sputtering to form $Nb_3Sn$ coating on a 2.6 GHz Nb SRF cavity using a custom-designed DC cylindrical magnetron sputtering system. By experimenting with various Nb and Sn composite ring targets arranged in tube configurations and placed on a cylindrical magnetron, optimal sputtering discharge conditions were determined to deposit films with Nb and Sn content suitable for $Nb_3Sn$ formation upon annealing. We report on the optimization of the co-sputtering deposition process, including adjustments to Nb-Sn content, coating thickness uniformity, and annealing conditions, along with an analysis of the RF performance of the $Nb_3Sn$-coated cavity. Finally, an additional light Sn recoating process was applied to the $Nb_3Sn$-coated cavity to improve the homogeneity of the film's surface, and the RF performance of the cavity after light Sn recoating was measured.

## 2. Experimental methods

### 2.1. Cylindrical sputter deposition system

The cylindrical magnetron sputtering system used for deposition has two magnetrons, developed by Plasmionique Inc., positioned axially at the top and the bottom of the sputtering chamber, with each containing four cylindrical magnets [31]. For co-sputtering deposition, only the bottom magnetron is used. Nb and Sn ring targets (0.9" OD × 0.8" ID) of varying lengths were fabricated from their original cylindrical targets (0.9" OD × 0.8" ID × 4.5" long, 99.99 % purity, from ACI Alloys Inc.) using electrical discharge machining (EDM). To eliminate surface contaminants, the targets underwent buffered chemical polishing (BCP) for the removal of about 100 μm of material from the target surface using a solution of HF (49 %), $HNO_3$ (70 %), and $H_3PO_4$ (85 %) in a 1:1:1 volume ratio.

The deposition process is controlled by a custom software (STP2C0) developed by Plasmionique Inc., which regulates both the sputtering discharge and the movement of the magnetron inside the cavity. The software raises the magnetron from its home position to the base of the cavity, initiates the discharge at a specified current, completes a preset number of passes across the cavity, and terminates the discharge. During both the discharge and non-discharge periods, a thermoelectric recirculating chiller, by ThermoTek model T257P, maintains water circulation through the magnetrons with a water temperature of 10 °C and with a flow rate of about 0.5 L/min through the magnetron to ensure effective cooling of the sputtering target.

## 2.2. Sputter target configuration

Nb and Sn ring targets are stacked around the cylindrical magnetron. Co-sputtering is carried out using this segmented composite target mounted on the bottom cylindrical magnetron, as shown in Fig. 1(a), in which Nb and Sn rings occupy different fractional areas of the composite target surface. During the magnetron discharge, distinct plasma glow rings are observed, as shown in Fig. 1(b). These plasma rings interact with the Nb and Sn target surfaces, resulting in the simultaneous sputtering of Nb and Sn, leading to mixed Nb-Sn atomic flux arriving at the substrate. Achieving the desired Nb and Sn composition in the as-deposited films requires optimization of the target configuration, discharge current, deposition pressure, Ar gas flow rate, and the speed at which the composite Nb-Sn target moves inside the SRF cavity.

The lengths and positions of the Nb and Sn ring targets are adjusted to optimize the Nb and Sn content in the sputtered film. The optimal co-sputtering target arrangement positioned on the bottom magnetron is shown in Fig. 1(a). The surface area of the magnetron covered by these targets is determined through co-sputtering depositions and analysis of the atomic composition in the as-deposited films. After a series of co-sputtering depositions aimed at optimizing the target configuration and achieving the desired film composition, the arrangement shown in Fig. 1(a) is identified as producing a composition suitable for forming $Nb_3Sn$ upon annealing. A spacer cap, used to secure the Nb and Sn targets, was fabricated from aluminum and coated with a about 1 μm sputtered Nb film. Figure 1(b) is a photograph showing emission from the Ar plasma generated when the magnetron is powered using 53 mA magnetron current with the Ar pressure 11 mTorr. The discharge shows enhanced glow intensity in the regions across the Nb and Sn targets, corresponding to the positions of the ring magnets inside the magnetron. Details of the magnet arrangement consisting of four magnets and the corresponding magnetic field distribution around the magnetron were published previously [31]. The plasma glow intensity across the composite Nb-Sn target is not uniform; this variation could be attributed to the higher secondary electron emission coefficient of Nb compared to Sn, which enhances local ionization efficiency and results in a higher plasma glow intensity in the Nb region compared to the Sn regions.

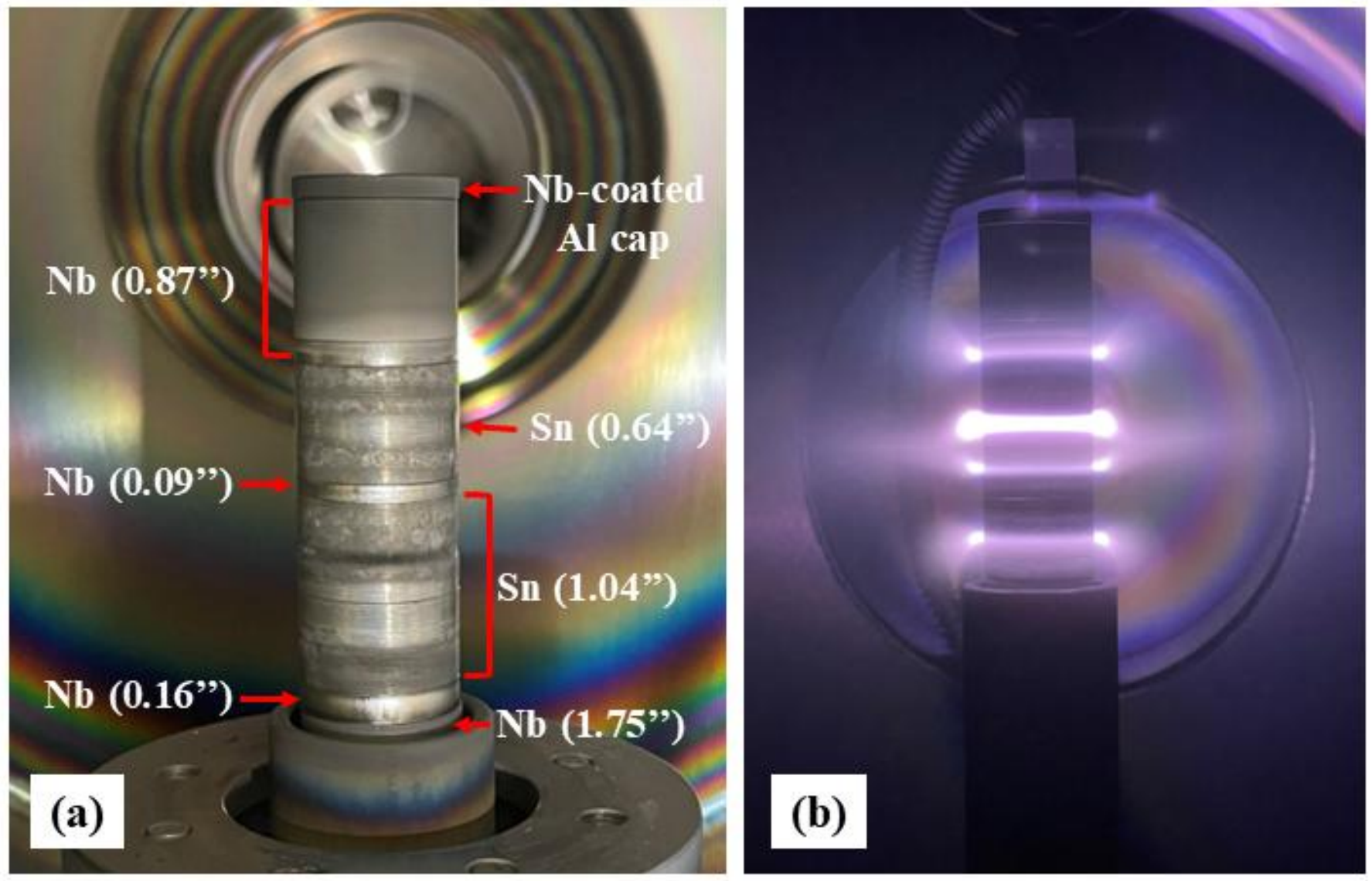

**Fig. 1.** Photograph of (a) the target configuration for optimized co-sputtering deposition using Nb and Sn ring targets on the cylindrical magnetron, including two Sn ring targets (of lengths 1.04", and 0.64") and one Nb ring target (of length 0.09"), that produce a Sn content in the as-deposited Nb-Sn samples that is suitable for forming $Nb_3Sn$ upon annealing. The 0.16", 0.87" Nb tubes on the bottom and top of the target configuration, respectively, are outside the location of the four magnets. (b) The cylindrical magnetron discharges are operated at 53 mA, generating an Ar plasma that exhibits enhanced glow intensity across the Nb and Sn targets in the regions adjacent to the magnets during co-sputter deposition.

## 2.3. Nb-Sn co-sputtered on flat substrates

An aluminum mockup cavity was made to replicate the deposition conditions of a 2.6 GHz Nb SRF cavity and to optimize the deposition conditions on small flat samples. The mockup cavity along with a sample holder was used to mount three Nb samples (10 mm × 10 mm × 3 mm) or Si wafers at the top beam tube, equator, and bottom beam tube. Using the optimal target configuration, shown in Fig. 1(a), with the magnetron current fixed at 53 mA, co-sputtering was carried out on Si substrates inside the mockup cavity. During the deposition inside the mockup cavity, adjustments were made to the magnetron speed to improve the thickness uniformity of the deposited film across the mockup cavity. The magnetron speed was set to 0.40, 0.13, and 0.30 mm/s across the top beam tube, equator, and bottom beam tube, respectively. Additionally, to improve uniformity of film thickness, the magnetron motion was paused at the equator position for 1 min during each pass. After the bottom magnetron completed 12 passes across the cavity during co-sputtering, starting at the lower end of the bottom beam tube and ending at the upper end of the top beam tube of the mockup cavity, the measured thicknesses of the Nb-Sn film on the Si samples mounted at the top beam tube, equator, and bottom beam tube were found to be about 486, 433, and 533 nm, respectively. This corresponds to a deposition rate of about 40, 36, and 44 nm per pass of the bottom magnetron across the mockup cavity. The co-sputtered Nb-Sn films on Si substrates from the top beam tube, equator, and bottom beam tube positions exhibited Sn contents of about 32, 38, and 29 at. %, respectively. During the deposition process, a cooling period of 40 min was included after every two passes of the magnetron discharge across the cavity. This time in which the discharge was turned off was added to allow the magnetron and Sn sputtering targets to cool down, thus preventing the Sn target from melting.

Following these deposition rates and conditions, Nb-Sn films were co-sputtered on Nb substrates. Flat Nb substrates were cut by EDM from high-purity Nb sheets (obtained from Tokyo Denkai Co., Japan) with a residual resistivity ratio (RRR) of about 300. Next, these samples were treated with BCP for the removal of about 100 μm of material from the surface. The substrates were sequentially cleaned by wiping with ethanol, then acetone, and finally ethanol again, and dried with dry air before mounting in the chamber. Using the mockup cavity, the flat Nb substrates were placed at three locations that replicate the top beam tube, equator, and bottom beam tube of a 2.6 GHz Nb SRF cavity. The baseline pressure of the chamber reached $3 \times 10^{-7}$ Torr, and the deposition pressure was maintained at 11 mTorr Ar at a flow rate of 50 SCCM. The bottom magnetron was operated at constant current mode with a discharge current of 53 mA (maintained DC power in the range of 17.2 – 18.2 W) to conduct the deposition on the flat Nb substrates. The bottom magnetron completed a total of 42 passes across the mockup cavity and deposited about 1.7, 1.5, and 1.8 μm Nb-Sn film on samples positioned at the top beam tube, equator, and bottom beam tube, respectively.

### 2.4. Deposition conditions on beam tubes and equator

Figure 2 presents the magnetron power profile during its movement across the cavity. Each discharge-on period consists of two halves: in the first half, the magnetron moves from the bottom to the top of the cavity, while in the second half, it moves back to its initial position. The inset picture shows the relative positions of the magnetron at key locations during the scan. The magnetron power increases as it moves through the bottom beam tube, reaching 17.8 W between positions 1 and 2, then decreases across the equator region, reaching the lowest value of 17.2 W between positions 2 and 3. The highest magnetron power of 18.2 W is observed when the top plasma ring reaches the upper end of the top beam tube of the mockup cavity (position 4), just before reversing its direction to return to the initial position. The magnetron current was fixed at 53 mA, while the voltage, and consequently, magnetron power, can vary depending on discharge conditions. A noticeable increase in the magnetron power is observed when the magnetron is inside the beam tubes, with the power reaching a peak when the magnetron is fully inside the beam tubes. The composite Nb-Sn target is about 8 and 40 mm away from the inside surface of the beam tubes and equator positions, respectively. The discharge voltage reaches a maximum of 343 V with the magnetron fully inside the beam tube and is a minimum of 324 V when the magnetron midpoint is at the equator position. At the surface of the target, the magnetic field strength, measured with a DC Gaussmeter (Model 1-ST, AlphaLab Inc.), varies between 1780 to 2900 G, while 8 mm away from the surface of the target it was between 130 to 245 G.  In a model for the DC planar magnetron discharge operated in Ar with a maximum horizontal magnetic field component at the target surface of 600 G, the cathode sheath thickness was found to decrease with the discharge voltage. At a voltage of 225 V and Ar pressure of about 10 mTorr, the sheath thickness was about 15 mm [32]. The increase in magnetron discharge power (and discharge voltage) when the magnetron is inside the beam tube is attributed to plasma-wall interactions, which could lead to the loss of electrons and ions through wall recombination. Since the discharge operates at a constant current, an increase in discharge power is necessary to compensate for the additional recombination reactions. This plasma-surface interaction is known to alter the film growth conditions causing mixing of the deposited Nb-Sn, some physical sputtering of the grown film, breaking surface bonds resulting in adsorption sites, and surface heating and electronic excitation enhancing surface diffusion [33,34]. Therefore, plasma-surface interaction at the beam tube regions can alter the deposition conditions from that at the equator region. Figure 2 (upper picture) shows the plasma discharge when the bottom magnetron reaches the upper end of the top beam tube during co-sputter deposition inside the mockup cavity. It is observed that the plasma extends to the top beam tube wall, indicating the presence of plasma-surface interactions, which is likely to influence the deposition conditions and resultant film characteristics in the beam tube region.

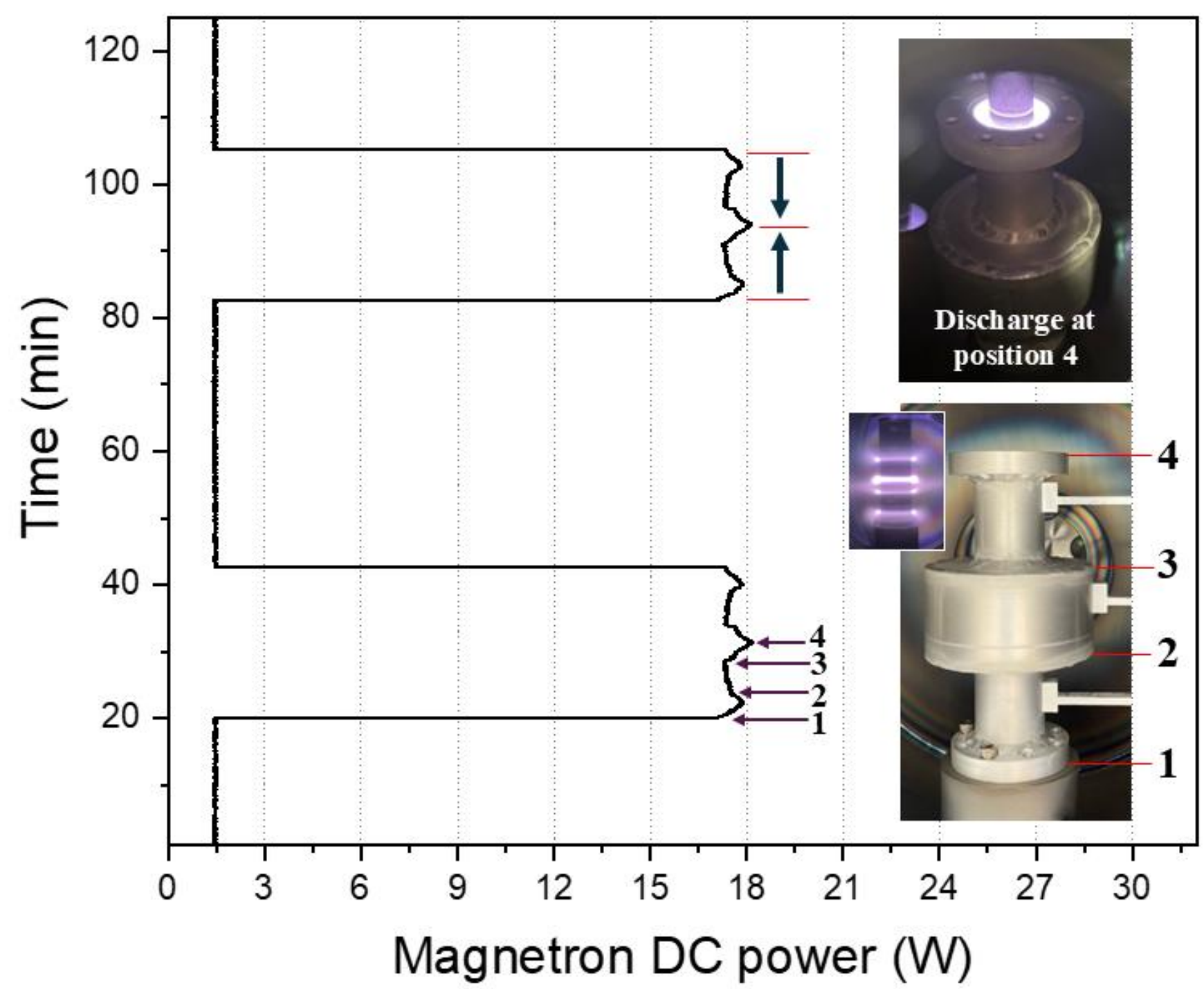

**Fig. 2.** Magnetron power during co-sputtering of the Nb-Sn film using a fixed magnetron discharge current of 53 mA. A 40-min cooling period with the discharge turned off was included after every two passes of the magnetron across the cavity. The numbers on the power trace correspond to the locations of the magnetron inside the mockup cavity. The top photograph shows the plasma discharge when the bottom magnetron reaches the upper end of the top beam tube of the mockup cavity (position 4) during co-sputtering, showing the plasma is filling the gap between the magnetron and the inside surface of the beam tube.

## 2.5. Annealing process

The sputtered samples were annealed in a high-vacuum furnace at Jefferson Lab to form the $Nb_3Sn$ phase by reacting Nb and Sn in the deposited Nb-Sn film. A detailed description of the furnace is available in [18]. A set of co-sputtered samples, those positioned at the mockup cavity's top beam tube, equator, and bottom beam tube, was placed inside a Nb sample chamber, and both ends were closed with Nb covers inside the clean room. Figure 3(a) shows Nb and sapphire samples coated with co-sputtered Nb-Sn films inside the sample chamber during the setup. The assembly was installed into the furnace insert for annealing, as shown in Fig. 3(b) during the loading process. The temperature was monitored with a type C thermocouple touching the sample chamber in selected experiments, see Figure 3(b). Annealing was initiated after the insert was pumped down to a low $10^{-5}$ Torr range. It typically consists of ramping the program temperature at a rate of 12 °C/min from room temperature to the annealing temperature of 950 ± 10 °C and holding it for 3 h. Using similar annealing conditions in the same furnace, previous studies have demonstrated that annealing of co-sputtered Nb-Sn films on Nb substrates at 950 $^0$C for 3 h produces high-quality, crystalline $Nb_3Sn$ films with densely packed grains and a $T_c$ of 17.61 K, as reported in [25].

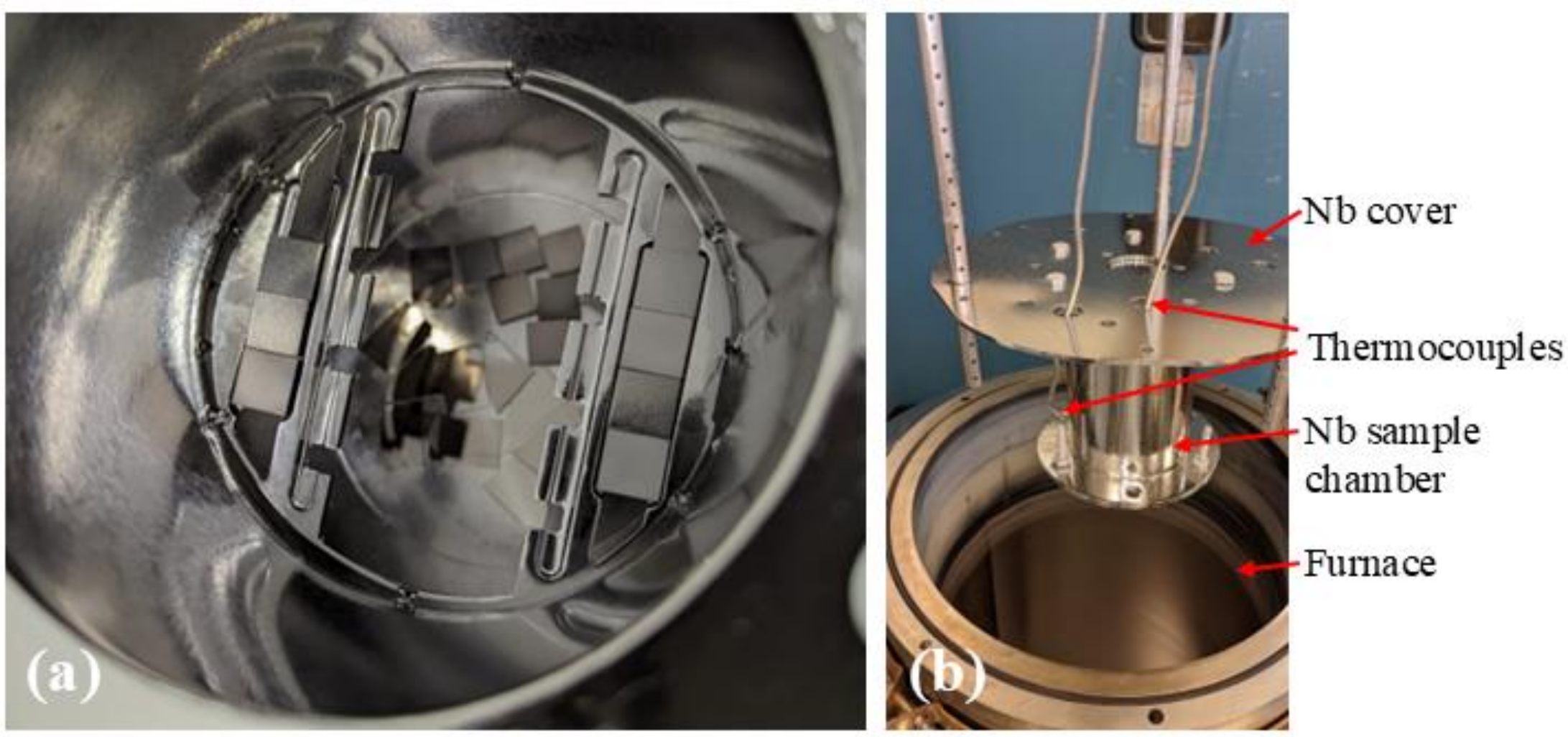

**Fig. 3.** Photograph of (a) Nb and sapphire samples coated with co-sputtered Nb-Sn films placed inside the sample chamber before the annealing and (b) loading the sample chamber with Nb samples into the furnace insert for an annealing treatment. Two thermocouples are placed in contact with the sample chamber, one at the top and one at the bottom, to monitor the temperature throughout the annealing process.

The microstructural analysis of annealed samples revealed significant differences between the equator and beam tube samples, with the equator samples exhibiting surface voids and lower Sn content. The detailed results are discussed in **Section 3.1**. Based on the surface morphology observed after annealing at 950 °C for 3 h, the annealing condition was systematically optimized to improve surface uniformity and reduce void density. Experiments were conducted by varying the duration of the 950 °C annealing and adding a preliminary annealing step at 600 °C before it for different periods. Table 1 provides a list, and details of the annealing conditions used in this study to enhance the surface quality of $Nb_3Sn$ coating.

**Table 1.** Samples prepared for the optimization study of annealing conditions to improve the surface quality of the $Nb_3Sn$ film by co-sputtering on Nb substrates.

| Annealing condition | Preliminary annealing | Duration of annealing at 950 °C | Sn content in the annealed samples (at. %) | | Void coverage on equator sample (% of total surface area) |
|---|---|---|---|---|---|
| | | | Top beam tube sample | Equator sample | |
| 1 | No | 3 h | 23 | 3 | 25 |
| 2 | No | 1 h | 23 | 13 | 28 |
| 3 | 600 °C – 1 h | 3 h | 23 | 11 | 16 |
| 4 | 600 °C – 3 h | 1 h | 22 | 11 | 23 |
| 5 | 600 °C – 4 h | 1 h | 23 | 13 | 26 |
| 6 | 600 °C – 6 h | 1 h | 22 | 10 | 9 |

## 2.6. Light Sn recoating of annealed samples

To improve the surface uniformity of the annealed samples, we explored an alternative approach involving light recoating using Sn vapor diffusion at Fermilab. This technique was previously implemented inside of a $Nb_3Sn$ coated 1.3 GHz Nb cavity and has been proven effective in healing the cracks on the $Nb_3Sn$ coating [35]. In this process, $Nb_3Sn$ coated cavity was annealed at 1000 °C, while Sn vapor was generated from a source held at 1250 °C, and the process was performed for 1 h. The amount of Sn used to produce the Sn vapor was 0.85 g. This light Sn recoating has been shown to significantly restore the performance of degraded $Nb_3Sn$ coated Nb cavity [35]. We applied light Sn recoating to reduce void density and enhance surface homogeneity in co-sputtered samples after annealing. Following the annealing process, light Sn recoating treatment was applied on $Nb_3Sn$ films deposited on flat Nb substrates and $Nb_3Sn$ coated 2.6 GHz Nb SRF cavity.

## 2.7. Film characterization

The crystal structure of the as-deposited Nb-Sn film and the annealed samples were examined by an X-ray diffractometer (XRD) (Rigaku Miniflex II, Japan) using Cu-Kα radiation (λ = 1.5406 Å) in the Bragg–Brentano geometry. The average crystallite size was estimated using the Scherrer equation given as Eq. (1) [36].

$$D = \frac{K\,\lambda}{\beta\cos\theta} \quad , \qquad (1)$$

where $D$ is the crystallite size, $K$ is the shape factor, $\lambda$ is the X-ray wavelength, $\beta$ is the full width at half maximum (FWHM) of the diffraction peak (in radians), and $\theta$ is the Bragg angle. The peak broadening attributed to microstructural effects was evaluated by correcting the measured FWHM for the instrumental response according to Eq. (2) [37].

$$\beta = (\beta_{measured^2} - \beta_{instrumental^2})^{1/2} \quad , \qquad (2)$$

where $\beta$, $\beta_{measured,}$ and $\beta_{instrumental}$ are the microstructural, observed, and instrumental FWHM, respectively. The lattice parameter $a$ of the $Nb_3Sn$ films was calculated using $Nb_3Sn$ (210) diffraction peak using the cubic crystal structure using Eq. (3) [36].

$$a = \frac{\lambda}{2\sin\theta}\sqrt{h^2 + k^2 + l^2} \quad , \qquad (3)$$

The elemental composition of as-deposited and annealed Nb-Sn films was analyzed using a Noran 6 energy dispersive X-ray spectroscopy (EDX) detector attached to a JOEL JSM 6060 LV SEM, with measurements conducted at an accelerating voltage of 15 kV. EDX analysis was performed on multiple surface areas for each sample, covering 1.2 square mm, and the mean composition was calculated. The surface morphology of as-deposited Nb-Sn films and annealed samples was analyzed using field emission scanning electron microscopy (FESEM S-4700, Hitachi, Japan). Surface topography was characterized using a Dimension Edge Atomic Force Microscope (AFM) from Bruker Corporation, operated in tapping mode.

## 3. Results and discussion

This section presents the analysis of the $Nb_3Sn$ layer formed upon annealing (950 °C for 3 h, referred to as annealing condition 1 in Table 1) of the Nb-Sn co-sputtered films on Nb substrates, prepared at the top beam tube, equator, and bottom beam of the mockup cavity. Next, the results of different annealing strategies (annealing conditions 2 – 6 in Table 1) used to optimize the annealing process for mitigating Sn loss and improving surface homogeneity are discussed. The cryogenic RF test results for the $Nb_3Sn$ coated 2.6 GHz Nb cavity, following annealing under the optimized condition, are then presented. Finally, this section concludes with the cryogenic RF test results after applying a light Sn recoating treatment further to enhance the performance of the co-sputtered $Nb_3Sn$ coated cavity.

### 3.1. Annealed films on flat substrates

### 3.1.1. Structure

Figure 4(a) shows the XRD patterns of the as-deposited Nb-Sn film co-sputtered on Nb substrates in locations representing the beam tubes and the equator, while Fig. 4(b) shows the XRD pattern of these samples after annealing at 950 °C for 3 h, referred to as annealing condition 1 in Table 1. As shown in Fig. 4(a), all the as-deposited samples exhibited Nb diffraction peaks (200), (211), and (310) contributed from the Nb substrates. In addition, the as-deposited samples from both the beam tubes and equator positions display a broad diffraction feature near the 2θ position of the $Nb_3Sn$ (210) reflection. This broad peak suggests the presence of a poorly crystalline Nb-Sn phase with short-range order, indicating the initial formation of $Nb_3Sn$. After annealing at 950 °C for 3 h, the $Nb_3Sn$ diffraction peaks (110), (200), (210), (211), (222), (320), (321), (400), (420), (421), and (332) were observed in all three annealed samples, as shown in Fig. 4(b). Annealed samples from all three positions show Nb diffraction peaks Nb (211), originating from the substrates. Annealed sample from the equator position demonstrated additional Nb diffraction peaks, Nb (110), (200), (220), and (310) in the diffraction pattern, contributed by the Nb substrate. No other diffraction peak from any Nb-Sn compound, apart from $Nb_3Sn$, was observed in the diffraction pattern of the annealed samples, indicating that only $Nb_3Sn$ formed after annealing.

The crystallite size of the annealed samples calculated using the $Nb_3Sn$ (210) diffraction peak was found to be about 145, 110, and 107 nm for the top beam tube, equator, and bottom beam tube samples, respectively. The lattice parameters of the annealed $Nb_3Sn$ films for the beam tubes and equator samples were found to be comparable, ranging from 0.517 to 0.523 nm, consistent with values previously reported for $Nb_3Sn$ films [15].

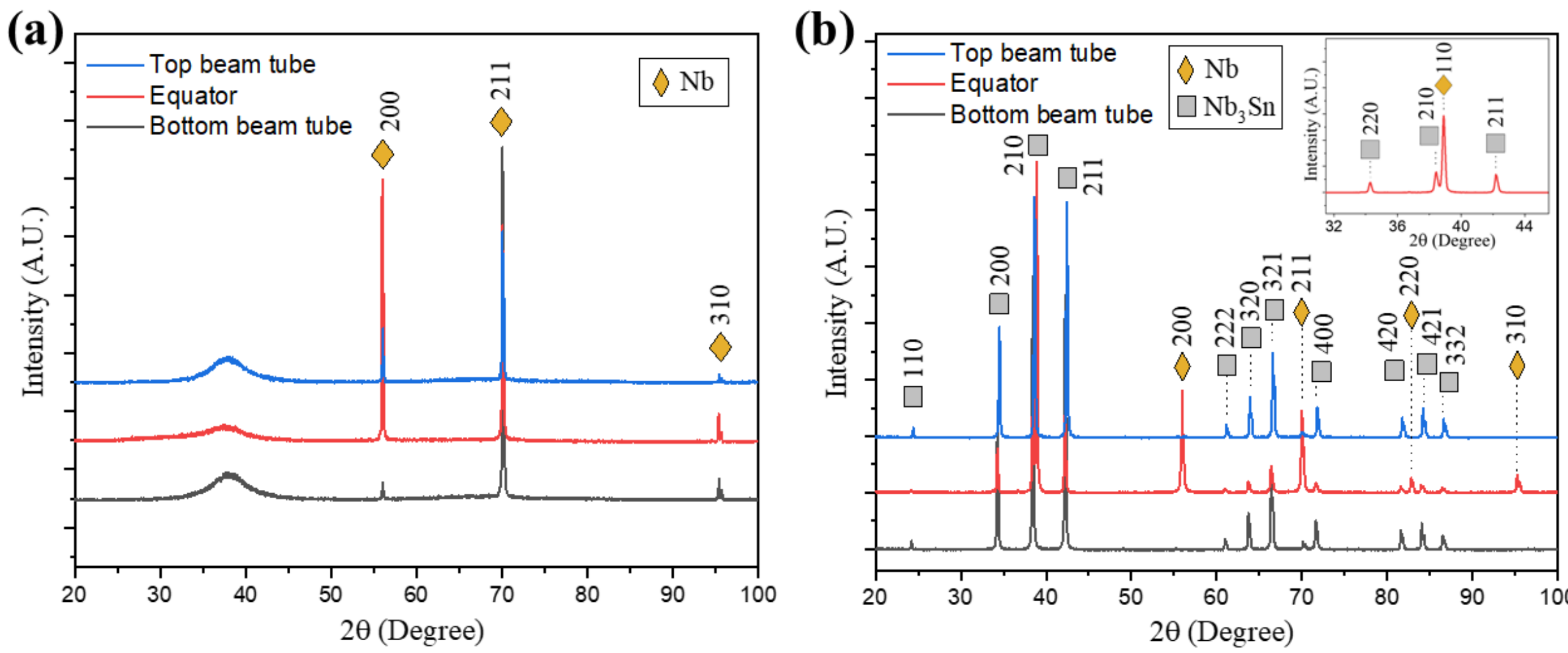

**Fig. 4.** XRD patterns of (a) as-deposited Nb-Sn films prepared by co-sputtering, and (b) films annealed at 950 °C for 3 h. The substrates were mounted in an Al mockup cavity at positions corresponding to the top beam tube, equator, and bottom beam tube, replicating a 2.6 GHz Nb SRF cavity.

### 3.1.2. Composition and surface morphology

Table 2 presents Sn content in the as-deposited Nb-Sn films and samples annealed at 950 °C for 3 h, referred to as annealing condition 1 in Table 1. The Sn content in the as-deposited top beam tube, equator, and bottom beam tube samples are 35, 42, and 32 at. %, respectively. For the plasma rings formed during the discharge, shown in Fig. 1(b), the Sn target tubes are positioned at the location of the top and bottom magnets, so the formed plasma rings from these two magnets sputter Sn. As the magnetron moves across the mockup cavity, deposition begins with the bottom Sn plasma ring near the cavity base, and by the end of the scan, the magnetron reaches the highest position of the cavity, where the top Sn plasma ring deposits material at the upper end of the top beam tube of the cavity. Consequently, the beam tube samples mostly receive Sn contributions from the top and bottom Sn plasma rings, respectively. In contrast, the equator sample, which receives contributions from all three Sn plasma rings during each pass of the bottom magnetron, exhibits a higher Sn content compared to the beam tube samples.

**Table 2.** Thickness and Sn content (at. %) of co-sputtered films in the as-deposited condition and after annealing at 950 °C for 3 h. The samples were prepared using the mockup cavity replicating the plasma geometry of a 2.6 GHz Nb SRF cavity, with Nb samples mounted at the beam tubes and the equator positions.

| Sample position | Co-sputtered film thickness (μm) | Sn content (at. %) in as-deposited samples | Sn content (at. %) in annealed samples |
|---|---|---|---|
| Top beam tube | 1.7 | 35 | 23 |
| Equator | 1.5 | 42 | 3 |
| Bottom beam tube | 1.8 | 32 | 23 |

The Sn content in the annealed samples is observed to be lower than the as-deposited samples. Sn evaporation resulting in subsequent reduction in Sn content during $Nb_3Sn$ growth was previously observed in co-sputtered Nb-Sn films after annealing [25]. As presented in Table 2, the annealed top beam tube and bottom beam tube samples both have a Sn atomic content of about 23 at. %, which falls within the expected range for $Nb_3Sn$. However, the annealed equator sample showed a significantly lower Sn content of only about 3 at. %.

Figure 5 shows the surface of the as-deposited films from the beam tubes and the equator positions of the mockup cavity, along with the films after annealing at 950 °C for 3 h (annealing condition 1 in Table 1). ImageJ surface analysis software was used to measure the size distribution of surface features. The surfaces of the as-deposited top and bottom beam tube samples, as shown in Fig. 5(a) and Fig. 5(c), respectively, exhibit randomly distributed surface structures with diameters ranging from about 16 to 34 nm, along with localized clustering in some regions. The as-deposited equator sample surface shown in Fig. 5(b) displays evenly distributed cracks in addition to finer surface features, primarily ranging from about 12 to 17 nm in diameter.

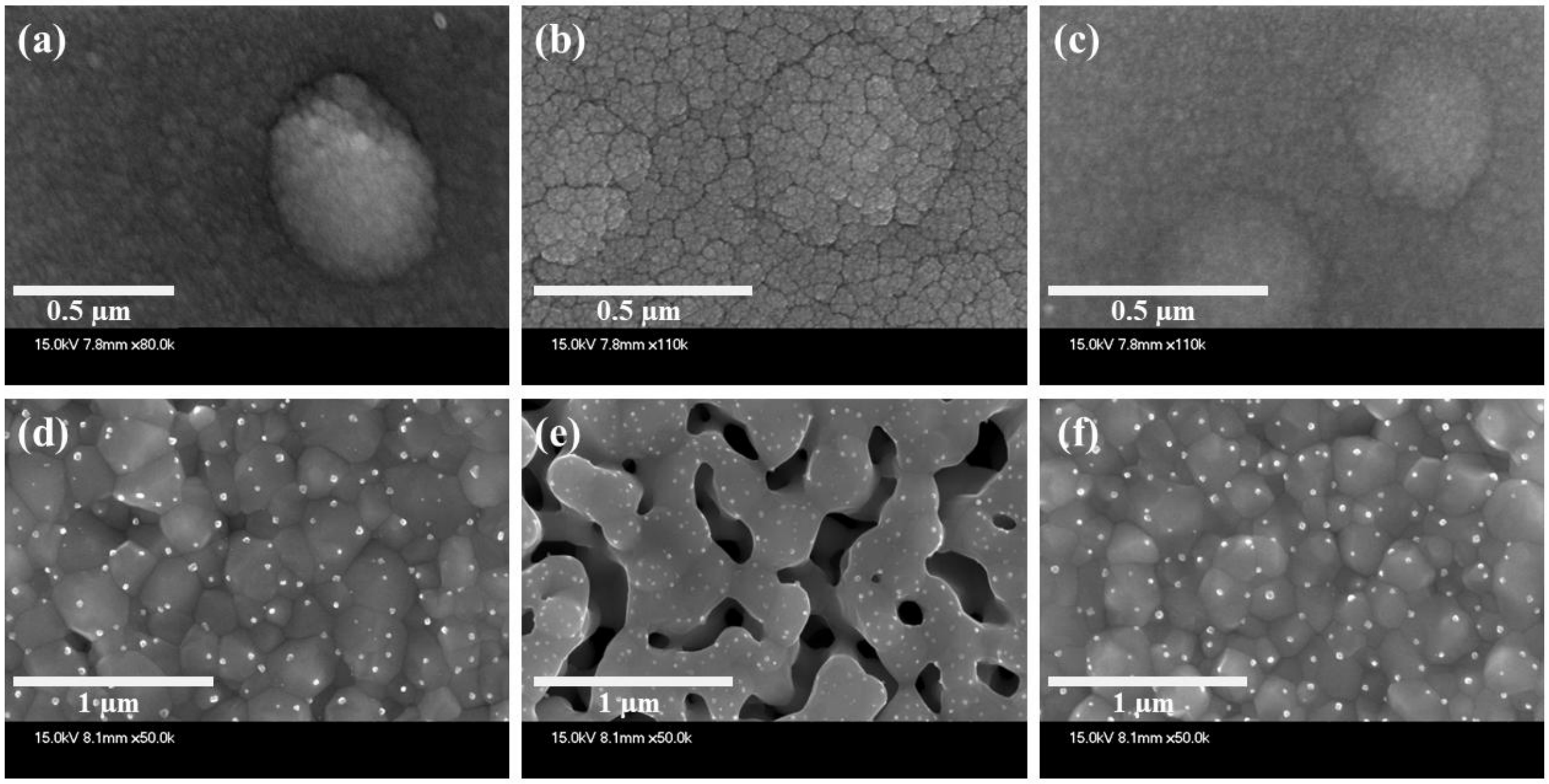

**Fig. 5.** FESEM micrographs of the film surface of as-deposited (a) top beam tube, (b) equator, and (c) bottom beam tube samples, prepared at equivalent positions of the 2.6 GHz Nb SRF cavity; (d), (e), and (f) are FESEM micrographs of film surfaces after annealing at 950 °C for 3 h of samples (a) – (c), respectively. The white dots visible on the annealed sample surfaces are Sn-rich particles that appear on both beam tubes and equator samples after annealing.

In contrast, the annealed sample surfaces from the top and bottom beam tubes, shown in Fig. 5(d) and Fig. 5(f), respectively, exhibit homogeneous and uniformly distributed $Nb_3Sn$ surface structures with clearly defined boundaries, consistent with a polycrystalline morphology with compact grain formation. The feature sizes on the annealed top beam tube sample range from about 147 to 374 nm in diameter, with an average of 264 nm, while those for the bottom beam tube sample range from about 140 to 392 nm, with an average of 254 nm. However, the annealed equator sample surface, shown in Fig. 5(e), exhibits a distinct morphology with elongated $Nb_3Sn$ structures lacking well-defined boundaries and the presence of large porous areas throughout the surface. The equator is located about 40 mm away from the surface of the sputter target, which is far from the magnetron plasma. However, the surface of the beam tubes is about 8 mm away from the sputter target, which allows the magnetron plasma to interact with the surface of the deposited Nb-Sn film. The plasma interaction involves energetic ion bombardment, causing the mixing of the deposited Nb-Sn, surface heating, enhanced adatom surface diffusion, and electronic excitation [33,34]. Figure 5(b) shows the development of surface cracks in the film after annealing. These cracks typically appear to relieve stress in the thin film [38]. The surface cracks observed on the as-deposited equator sample, shown in Fig. 5(b), may provide additional pathways for the rapid evaporation of Sn from the thin film layer, leading to increased Sn loss. This rapid evaporation resulted in a Sn content of about 3 at. % in the annealed equator sample, compared to about 23 at. % Sn in the annealed beam tube samples, likely contributing to the formation of the voids observed on the surface of the annealed equator sample, shown in Fig. 5(e). Additionally, small particle-like structures (18 – 36 nm in size), see bright features in Fig. 5(d, e, and f), are observed on the surfaces of all annealed samples. Comparing the composition of the annealed sample's surface containing these particles with the particle-free regions indicates that the particles are enriched with Sn content.

### 3.2. Optimizing annealing parameters

As shown in Fig. 5(b), the as-deposited equator sample surface exhibits cracks that, after annealing at 950 °C for 3 hours, develop into large void channels, along with elongated $Nb_3Sn$ grains, as shown in Fig. 5(e). To improve surface morphology and mitigate void formation on the equator sample surface upon annealing, we explored a few alternative strategies for annealing the co-sputtered samples. For the annealing optimization study, a series of co-sputtered samples were prepared inside the mockup cavity using the target configuration illustrated in Fig. 1(a) and the deposition conditions detailed in **Section 2.4**. The top beam tube and equator samples were deposited on Nb substrates to investigate surface properties upon annealing, while the bottom beam tube sample was deposited on a sapphire substrate (430 μm thick, double-side polished, C-plane orientation, from University Wafer Inc.) to assess the consistency of the film composition.

To examine the effect of high-temperature annealing duration, one set of co-sputtered samples on Nb substrates (top beam tube and equator sample) underwent annealing at 950 °C for 1 h. Another set of samples was subjected to a preliminary low-temperature annealing step at 600 °C for 1 h, followed by high-temperature annealing at 950 °C for 3 h. These annealing conditions are referred to as annealing conditions 2 and 3 in Table 1. Surface analysis revealed that reducing the high-temperature annealing duration from 3 h to 1 h and introducing a preliminary annealing step at 600 °C increased the Sn content in the annealed equator samples. Additionally, the equator sample that underwent preliminary annealing at 600 °C for 1 h (annealing condition 3) exhibited lower void density on the film surface. To further investigate these trends, the influence of varying the preliminary annealing duration at 600 °C, followed by a final annealing at 950 °C for 1 h, was examined. Three sets of co-sputtered samples, deposited on Nb substrates positioned at the top beam tube and equator, underwent preliminary annealing at 600 °C for 3, 4, and 6 h, followed by final annealing at 950 °C for 1 h, referred to as annealing conditions 4, 5, and 6 in Table 1.

### 3.2.1 Composition and morphology of annealed equator samples

The equator sample annealed at 950 °C for 1 h (annealing condition 2) exhibited a Sn content of about 13 at. %, while the equator sample that underwent preliminary annealing at 600 °C for 1 h, followed by annealing at 950 °C for 3 h (annealing condition 3), had a Sn content of about 11 at. %. These results indicate that both approaches reduced Sn loss compared to the equator sample directly annealed at 950 °C for 3 h (annealing condition 1), which resulted in only about 3 at. % Sn. Equator samples that underwent preliminary annealing at 600 °C for 3, 4, and 6 h, followed by 950 °C for 1 h (annealing conditions 4, 5, and 6, respectively), exhibited Sn contents of about 11, 13, and 10 at. %, respectively.

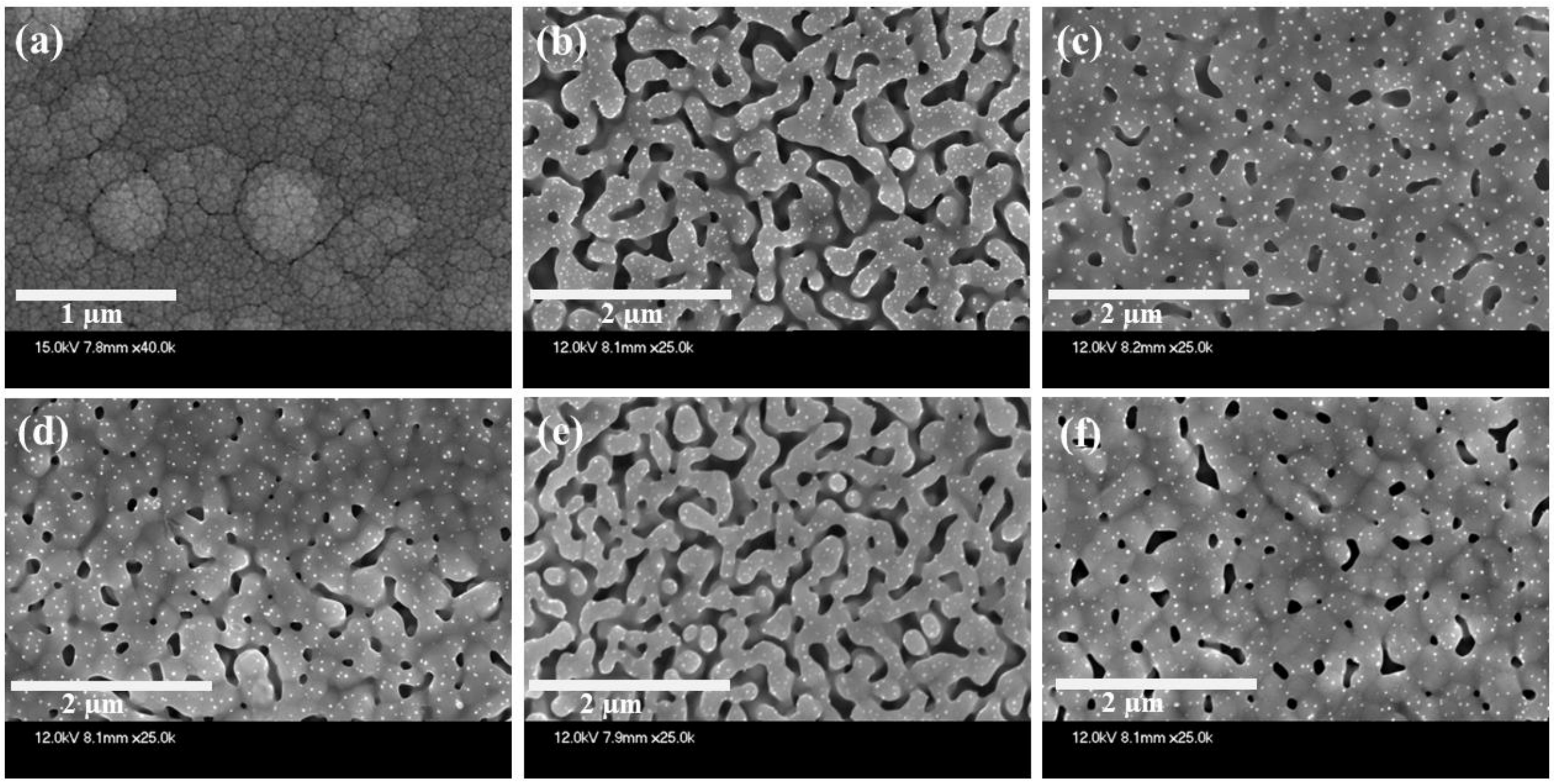

**Fig. 6.** FESEM micrographs of equator sample surfaces (a) as-deposited, and after annealing at (b) 950 °C for 1 h, (c) preliminary annealing at 600 °C for 1 h followed by 950 °C for 3 h, and after preliminary annealing at 600 °C for (d) 3 h, (e) 4 h, and (f) 6 h followed by annealing at 950 °C for 1 h. Annealing condition with preliminary annealing at 600 °C for 6 h, followed by annealing at 950 °C for 1 h, is found to be the most optimal condition, resulting in minimal void density and enhanced Sn content.

Figure 6(a) shows FESEM images of the as-deposited equator sample surface, showing microcracks and representing equator samples prepared in different deposition runs. Figures 6(b – f) show the annealed equator sample surfaces following annealing conditions 2 – 6, respectively, as detailed in Table 1. The equator sample annealed at 950 °C for 1 h (annealing conditions 2) and that received preliminary annealing at 600 °C for 1 h followed by 950 °C for 3 h (annealing conditions 3), both exhibited void formation on the annealed surface, as shown in Fig. 6(b – c). Quantitative analysis of void coverage was performed on 5 μm × 3.2 μm regions of the FESEM images using ImageJ, an image processing software, where contrast-based segmentation was applied to identify void regions, and the total void area was normalized to the total image area to determine the percentage void surface coverage. Equator sample annealed at 950 °C for 1 h (annealing condition 2) exhibited a void coverage of 28%, as shown in Fig. 6(b), while the sample that underwent preliminary annealing at 600 °C for 1 h, followed by annealing at 950 °C for 3 h (annealing condition 3), showed a reduced void coverage to 16%, as shown in Fig. 6(c). These results indicate that introducing a preliminary annealing step at 600 °C for 1 h before annealing at 950 °C for 3 h effectively reduced void density compared to the equator sample directly annealed at 950 °C for 3 h (annealing Condition 1), which exhibited 25% void coverage, as shown Fig. 5(e). FESEM images in Fig. 6(d – f) show the equator samples that underwent preliminary annealing at 600 °C for 3, 4, and 6 h, respectively, followed by final annealing at 950 °C for 1 h (annealing conditions 4, 5, and 6, respectively). The void surface area coverage on the equator samples subjected to these annealing conditions was 23, 26, and 9%, respectively. These findings demonstrate that the lowest void density is observed with preliminary annealing at 600 °C for 6 h followed by 950 °C for 1 h (annealing condition 6), achieving the most optimized annealed surface among the tested conditions. Furthermore, Sn-rich particles ranging between 12 – 36 nm in diameter were consistently observed on the annealed equator sample surfaces across annealing conditions 2 – 6, as shown in Fig. 6(b – f). The preliminary annealing step at 600 °C likely promotes solid-state interdiffusion and chemical homogenization of the Nb-Sn film, resulting in a more uniform distribution of Sn and mitigating rapid Sn depletion and void formation during subsequent high-temperature annealing at 950 °C for $Nb_3Sn$ growth.

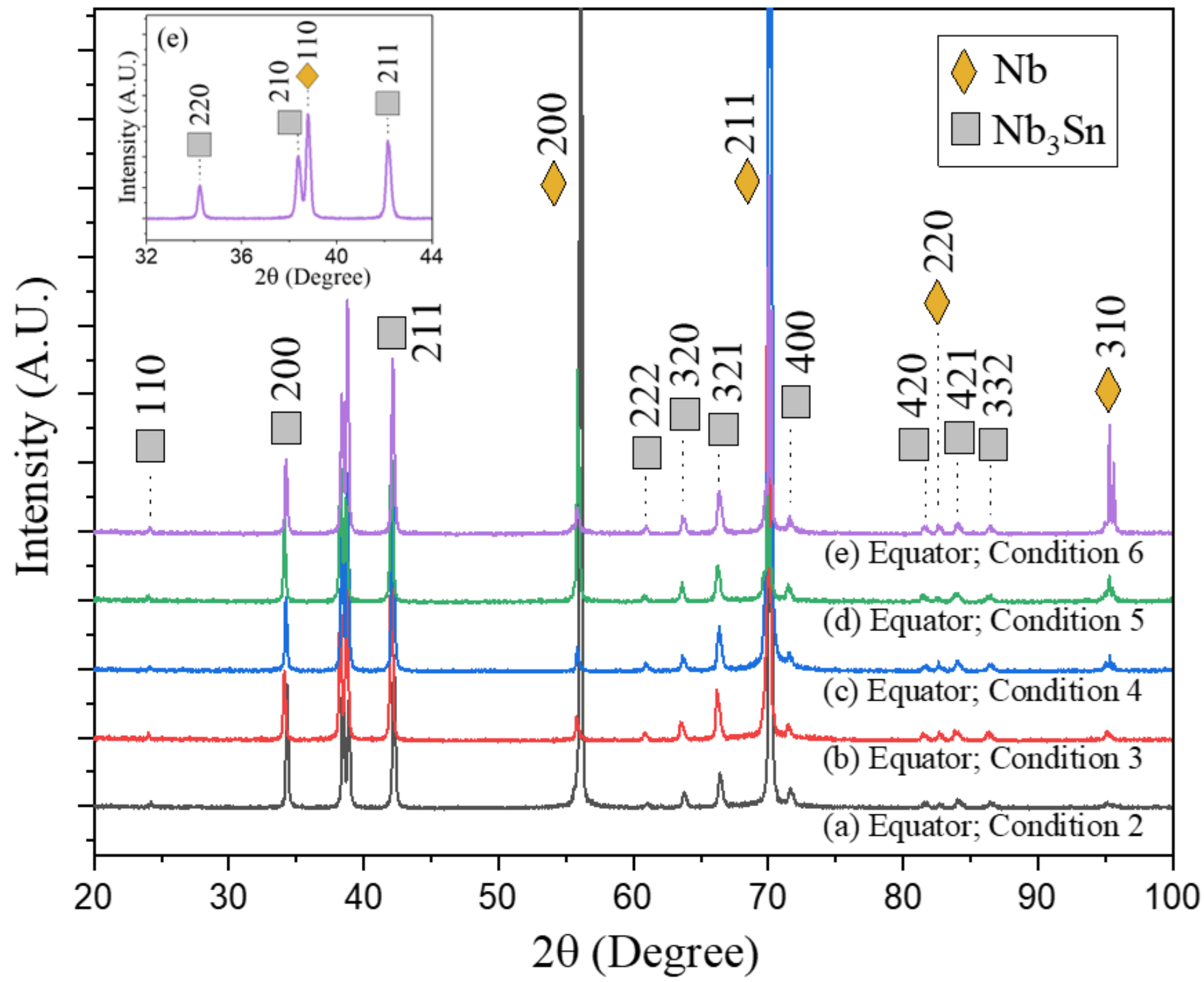

**Fig. 7.** XRD patterns of the equator samples after annealing at (a) 950 °C for 1 h, (b) 600 °C for 1 h followed by 950 °C for 3 h, and preliminary annealing at 600 °C for (c) 3 h, (d) 4 h, and (e) 6 h followed by annealing at 950 °C for 1 h. XRD results demonstrate that equator samples annealed under different conditions exhibit single-phase $Nb_3Sn$ diffraction peaks, with no other Nb-Sn compounds detected, confirming the formation of $Nb_3Sn$ only.

Figure 7 presents the XRD patterns of equator samples annealed under annealing conditions 2 – 6, as detailed in Table 1. All five annealed equator samples exhibited $Nb_3Sn$ diffraction peaks corresponding to (110), (200), (210), (211), (222), (320), (321), (400), (420), (421), and (332) planes. No diffraction peak from any other Nb-Sn compounds was observed, indicating that only $Nb_3Sn$ formed after annealing. Nb diffraction peaks corresponding to (110), (200), (211), (220), and (310) planes were observed in all annealed equator samples originating from the Nb substrate. The intensity of the Nb (310) peak increased with longer preliminary annealing durations, with the equator sample annealed under condition 6 (600 °C for 6 h followed by 950 °C for 1 h) exhibiting the strongest (310) peak. Additionally, Nb (200) peak is more pronounced for equator samples annealed under conditions 2 and 5 (950 °C for 1 h, and 600 °C for 4 h followed by 950 °C for 1 h, respectively), consistent with the higher surface void coverage observed in equator sample annealed under these two conditions as shown Fig. 6(a) and Fig. 6(d), respectively.

### 3.2.2. Composition and morphology of annealed top beam tube samples

The top beam tube samples annealed at 950 °C for 1 h (annealing condition 2) and preliminary annealed at 600 °C for 1 h, followed by annealing at 950 °C for 3 h (annealing condition 3), exhibited a Sn content of 23 at. %. These values are consistent with the Sn content of 23 at. % observed in the top beam tube sample annealed at 950 °C for 3 h (annealing condition 1). Similarly, the top beam tube samples subjected to preliminary annealing at 600 °C for 3, 4, and 6 h, followed by final annealing at 950 °C for 1 h (annealing conditions 4, 5, and 6, respectively), exhibited Sn contents of 22, 23, and 22 at. %, respectively. These results demonstrate that reducing the high temperature (950 °C) annealing duration or varying the preliminary annealing duration at 600 °C has minimal impact on the Sn content of the top beam tube samples.

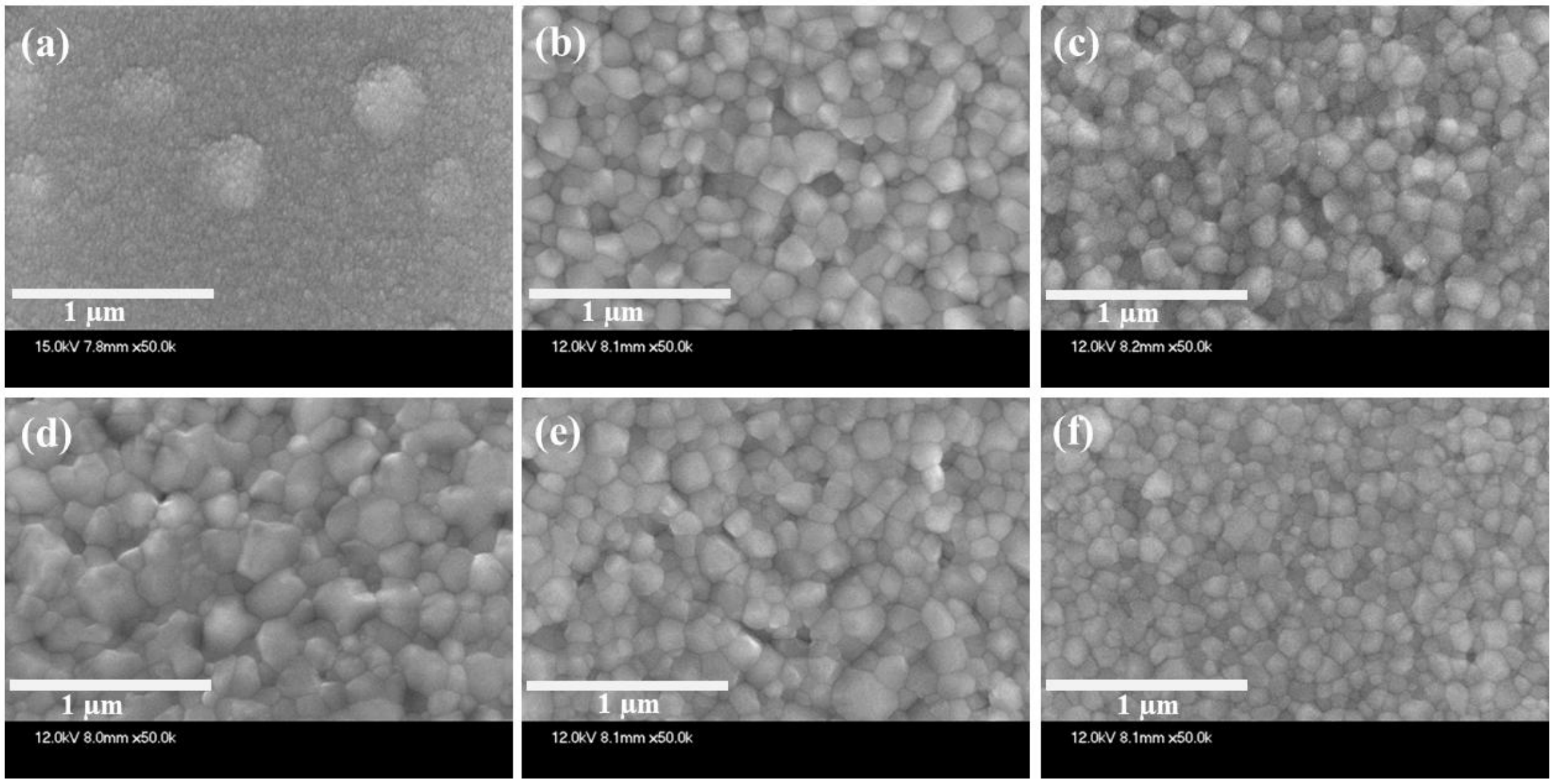

**Fig. 8.** FESEM micrographs of top beam tube samples (a) as-deposited, and after annealing at (b) 950 °C for 1 h, (c) 600 °C for 1 h followed by 950 °C for 3 h, and after preliminary annealing at 600 °C for (d) 3 h, (e) 4 h, and (f) 6 h followed by annealing at 950 °C for 1 h. All conditions result in $Nb_3Sn$ grain formation with well-defined grain boundaries. Increasing the duration of the preliminary annealing step tends to produce smaller $Nb_3Sn$ grains on the annealed top beam tube sample surface.

Figure 8(a) shows the as-deposited top beam tube sample surface with randomly distributed grains, serving as a reference for top beam tube samples produced across depositions. Figure 8(b – f) presents FESEM images of the surfaces of top beam tube samples annealed under conditions 2 – 6, respectively, as detailed in Table 1. These surface images exhibit distinct microstructural differences from the equator sample surfaces annealed under the same conditions, as shown in Figure 6(b – f). For annealing conditions 2 – 6, the top beam tube samples exhibit well-defined $Nb_3Sn$ grains with clear grain boundaries. Samples annealed at 950 °C for 1 hour (annealing condition 2) and those preliminarily annealed at 600 °C for 1 hour, followed by 950 °C for 3 hours (annealing condition 3), exhibit average $Nb_3Sn$ grain sizes of 149 nm and 152 nm, respectively, as shown in Fig. 8(b) and Fig. 8(c). These values are smaller than the average grain size of 264 nm for the top beam tube sample annealed at 950 °C for 3 h (annealing condition 1), presented in Fig. 5(a). Additionally, for the top beam tube samples $Nb_3Sn$ grain size decreases as the duration of preliminary annealing at 600 °C increases, measuring average grain sizes of 202, 178, and 137 nm for 3, 4, and 6 h preliminary annealed samples, respectively, as shown in Fig. 8(d – f), after final annealing at 950 °C for 1 h (annealing conditions 4, 5, and 6) was completed.

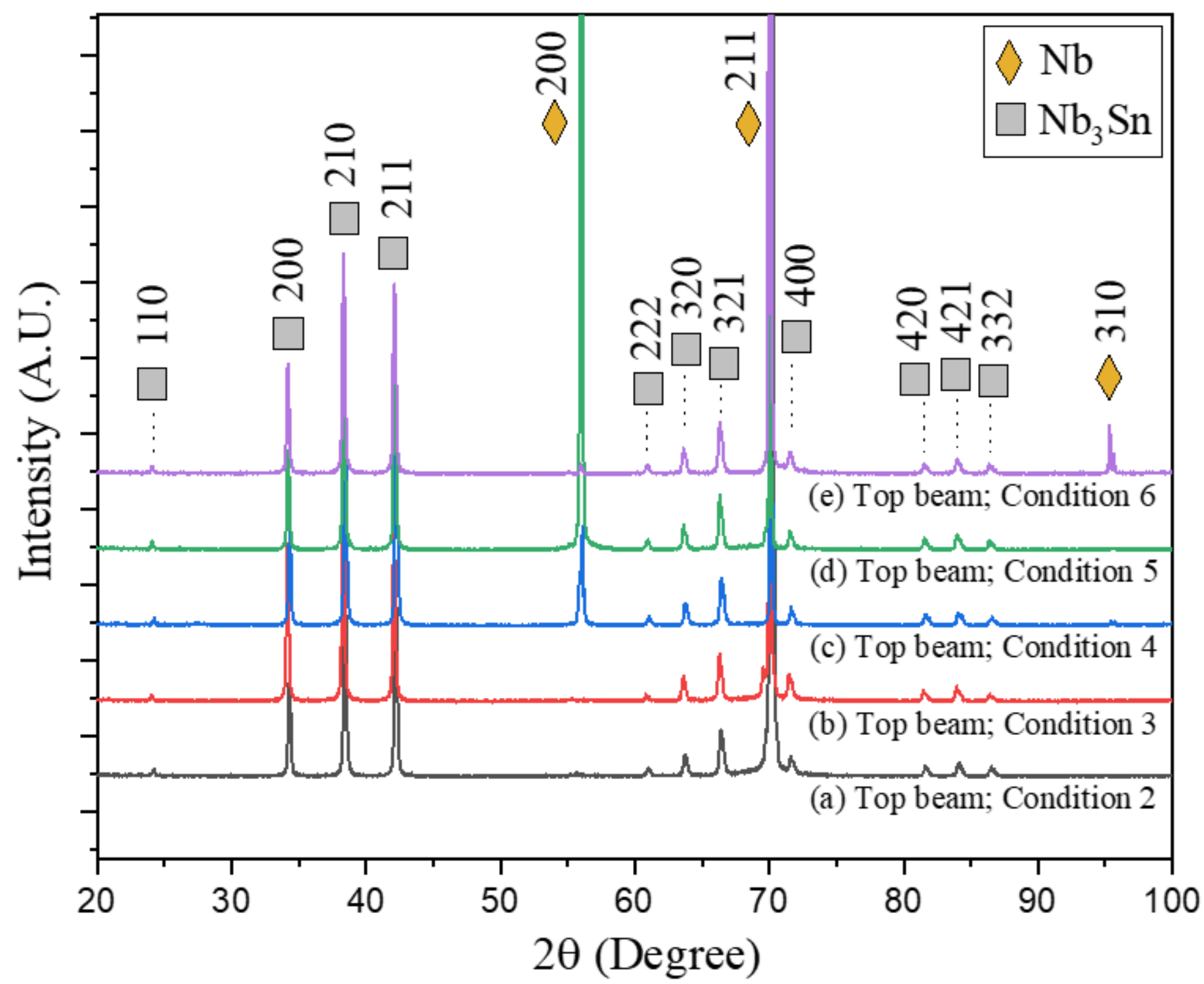

**Fig. 9.** XRD patterns of the top beam tube samples after annealing at (a) 950 °C for 1 h, (b) 600 °C for 1 h followed by 950 °C for 3 h, and after preliminary annealing at 600 °C for (c) 3 h, (d) 4 h, and (e) 6 h followed by annealing at 950 °C for 1 h. XRD results exhibit that top beam tube samples annealed under different conditions display single-phase $Nb_3Sn$ diffraction peaks, with no detectable peak from other Nb-Sn compounds.

Figure 9 displays the XRD patterns of top beam tube samples annealed under conditions 2 – 6, as outlined in Table 1. Similar to the XRD pattern of the annealed equator samples discussed in **Section 3.2.1**, all five annealed top beam tube samples exhibited $Nb_3Sn$ diffraction peaks corresponding to the

(110), (200), (210), (211), (222), (320), (321), (400), (420), (421), and (332) planes. Apart from $Nb_3Sn$, no diffraction peaks corresponding to any Nb-Sn compounds were observed. Nb diffraction peaks corresponding to the (211) planes, originated by the substrates, were observed in all annealed top beam tube samples. Samples annealed under conditions 4, 5, and 6 (preliminary annealed at 600 °C for 3, 4, and 6 h, respectively, followed by annealing at 950 °C for 1 hour) exhibited a Nb (200) diffraction peak from the substrate. Only the top beam tube sample annealed under condition 6 displayed an additional Nb (310) diffraction peak from the substrate.

### 3.3. Sn recoated samples

Four sets of co-sputtered samples, prepared on Nb substrates mounted at the top beam tube and equator positions of a mockup cavity, were treated with a light Sn recoating process [35]. Before Sn recoating, one set of samples was annealed at 950 °C for 1 h, while the remaining three sets underwent preliminary annealing at 600 °C for 3, 4, and 6 h, followed by annealing at 950 °C for 1 h.

The EDX analysis of the light Sn-recoated samples exhibited Sn content of 23 at. % in the equator samples. These results indicate a significant increase in Sn content in the equator samples after Sn recoating, compared to those that underwent similar annealing treatments without Sn recoating, detailed as annealing conditions 2, 4, 5, and 6 in Table 1. Following the light Sn recoating, the Sn content in the equator samples falls within 17 – 26 at. % range required for $Nb_3Sn$ formation, as defined by the Nb-Sn phase diagram [15]. The top beam tube samples exhibited 24 at. % Sn after the light Sn recoating process, which is close to the Sn content (22 – 23 at. %) of the top beam tube samples after annealing under conditions 2 – 6, as detailed in Table 1.

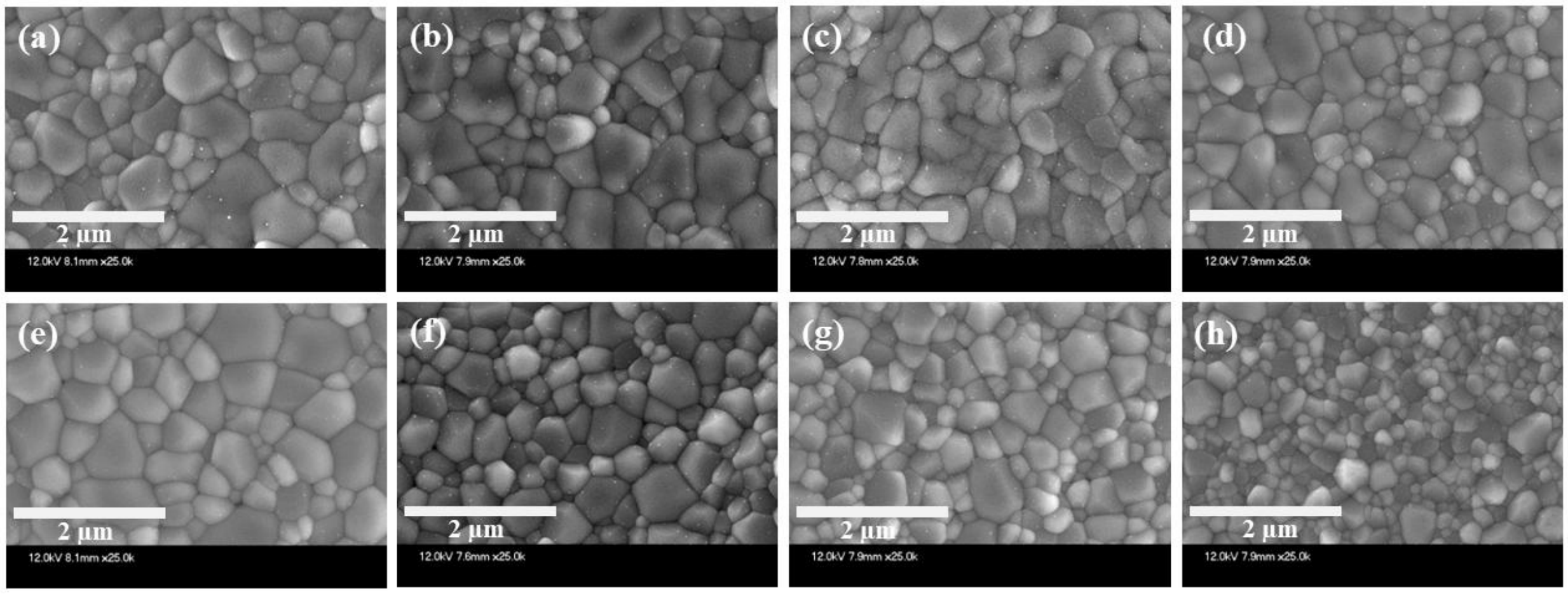

**Fig. 10.** FESEM micrographs of the equator samples after the light Sn recoating. Prior to the Sn recoating, the samples underwent annealing at (a) 950 °C for 1 h, and at 600 °C for (b) 3 h, (c) 4 h, and (d) 6 h, followed by annealing at 950 °C for 1 h. Corresponding FESEM images of top beam tube samples after the Sn recoating are shown in (e), (f), (g), and (h), which underwent the same annealing conditions as (a), (b), (c) and (d), respectively. Sn recoating treatment effectively eliminates surface voids on the equator samples and promotes the growth of larger $Nb_3Sn$ grains. Similarly, the top beam tube samples exhibit increased grain size following Sn recoating.

From the FESEM images shown in Fig. 10(a – d), Sn recoating treatment eliminated surface voids on the equator sample surface and formed $Nb_3Sn$ grains. It is speculated that Sn diffuses into the Nb substrate through the regions where voids previously existed, facilitating the formation of $Nb_3Sn$ in these areas and effectively eliminating the voids. After the light Sn recoating, large $Nb_3Sn$ grains were observed on the equator sample surface under various annealing conditions. Grain sizes reached up to 975 nm in length for the sample that underwent preliminary annealing at 600 °C for 6 hours, followed by annealing at 950 °C for 1 hour, as shown in Fig. 10(c).

After the Sn recoating treatment, all four top beam tube samples exhibited well-defined $Nb_3Sn$ grains with well-defined grain boundaries, as shown in Fig. 10(e – h). The top beam tube sample annealed at 950 °C for 1 h prior to Sn recoating had the largest average grain size of 585 nm in diameter, as shown in Fig. 10(e). The top beam tube samples that underwent preliminary annealing at 600 °C for 3, 4, and 6 h, followed by 950 °C annealing for 1 h, displayed progressively smaller grain sizes of 483, 455, and 343 nm in diameter, respectively, as shown in Fig. 10(f – h), indicating that, for the conditions studied, increased preliminary annealing duration reduces the $Nb_3Sn$ grain size. Surface topography analysis by AFM after Sn recoating on the $Nb_3Sn$ coated samples measured root-mean-square (RMS) roughness values ranges 18 – 25 nm for the equator samples and 20 – 25 nm for the top beam tube samples.

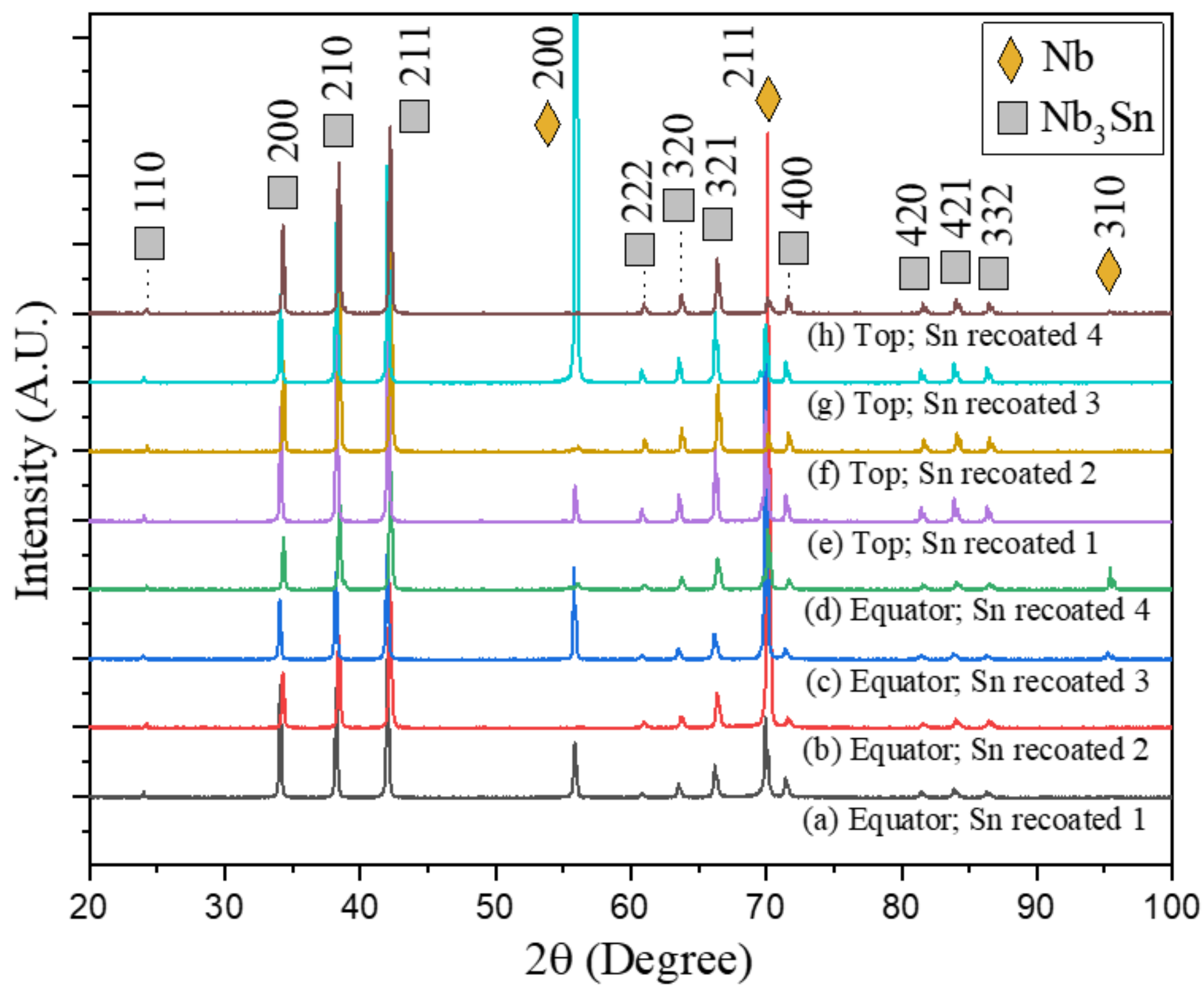

**Fig. 11.** XRD patterns of the equator samples after light Sn recoating. Prior to Sn recoating, the samples underwent annealing at (a) 950 °C for 1 h, and at 600 °C for (b) 3 h, (c) 4 h, and (d) 6 h, followed by annealing at 950 °C for 1 h. Corresponding XRD patterns of the top beam tube samples after light Sn recoating are shown in (e), (f), (g), and (h), which underwent the same annealing conditions as (a), (b), (c) and (d), respectively. XRD results show that after Sn recoating, both the top beam tube and equator samples exhibit similar $Nb_3Sn$ diffraction peaks, with the top beam tube samples showing stronger $Nb_3Sn$ diffraction peaks compared to the equator samples under the same annealing conditions.

Figure 11 presents the XRD patterns of Sn-recoated samples. The samples annealed with a single step annealing at 950 °C for 1 h, and those annealed at 600 °C for 3, 4, and 6 h followed by annealing at 950 °C for 1 h, are labeled as Sn-recoated 1 – 4, respectively. All four sets, for both the top beam tube and equator samples, display similar XRD peaks corresponding to $Nb_3Sn$ (110), (200), (210), (211), (222), (320), (321), (400), (420), (421), and (332) planes. The intensity of the $Nb_3Sn$ diffraction peaks is comparable to the top beam tube and equator samples; however, the $Nb_3Sn$ (420), (421), and (332) peaks are stronger in the top beam tube samples compared to the equator samples from the same deposition run. All Sn-recoated samples exhibit diffraction peaks corresponding to Nb (211) contributed by the Nb substrates. Both the top beam tube and equator samples annealed at 950 °C for 1 h, then Sn recoated and annealed at 600 °C for 4 h, followed by 950 °C for 1 h, then Sn recoated exhibit an additional Nb diffraction peak for Nb (200), also attributed to the substrate. Additionally, the equator samples annealed at 600 °C for 4 and 6 h, respectively, followed by 950 °C for 1 h, then Sn recoated display another Nb peak at Nb (310). Aside from the $Nb_3Sn$ and Nb diffraction peaks from substrate, no other diffraction peaks corresponding to any Nb-Sn compounds were observed.

### 3.4. SRF cavity coating by co-sputtering

A TESLA-shaped single-cell 2.6 GHz Nb SRF cavity made from high-purity Nb with a residual resistivity ratio (RRR) of about 300 was coated by $Nb_3Sn$ by co-sputtering of Nb and Sn. The cavity underwent a series of standard processing steps before installation in the cylindrical magnetron sputtering chamber for $Nb_3Sn$ coating [39]. Prior to sputtering, the cavity was anodized at 30 V. The processed cavity was cleaned and assembled in an ISO4 cleanroom for $Nb_3Sn$ coating in the cylindrical magnetron sputtering system.

The 2.6 GHz Nb SRF cavity was centered in the cylindrical magnetron sputtering system, as shown in the inset of Fig. 12. The bottom magnetron was used for co-sputtering, and the top magnetron had a Nb tube target installed. The same target arrangement, illustrated in Fig. 1(a), and the same automation program and deposition conditions used for co-sputtering deposition on flat Nb samples, detailed in **Section 2.3,** were used. The base pressure of the deposition chamber reached $3.7 \times 10^{-7}$ Torr. Before initiating the co-sputtering inside the cavity, the Nb tube target on the top magnetron was discharged at 30 W DC power for 5 min at its home position outside the SRF cavity, which improved the chamber vacuum from $3.7 \times 10^{-7}$ Torr to $1.9 \times 10^{-7}$ Torr via Nb gettering. During Nb/Sn co-sputtering, the bottom magnetron was operated in constant current mode at a discharge current of 47 mA and completed a total of 42 scans across the cavity. Figure 12 presents the magnetron power profile during cavity coating, showing that the discharge power was 16.7 – 17.6 W at a magnetron discharge current of 47 mA. The discharge was turned off for 80 minutes each time the magnetron returned to its home position to avoid overheating the sputtering target.

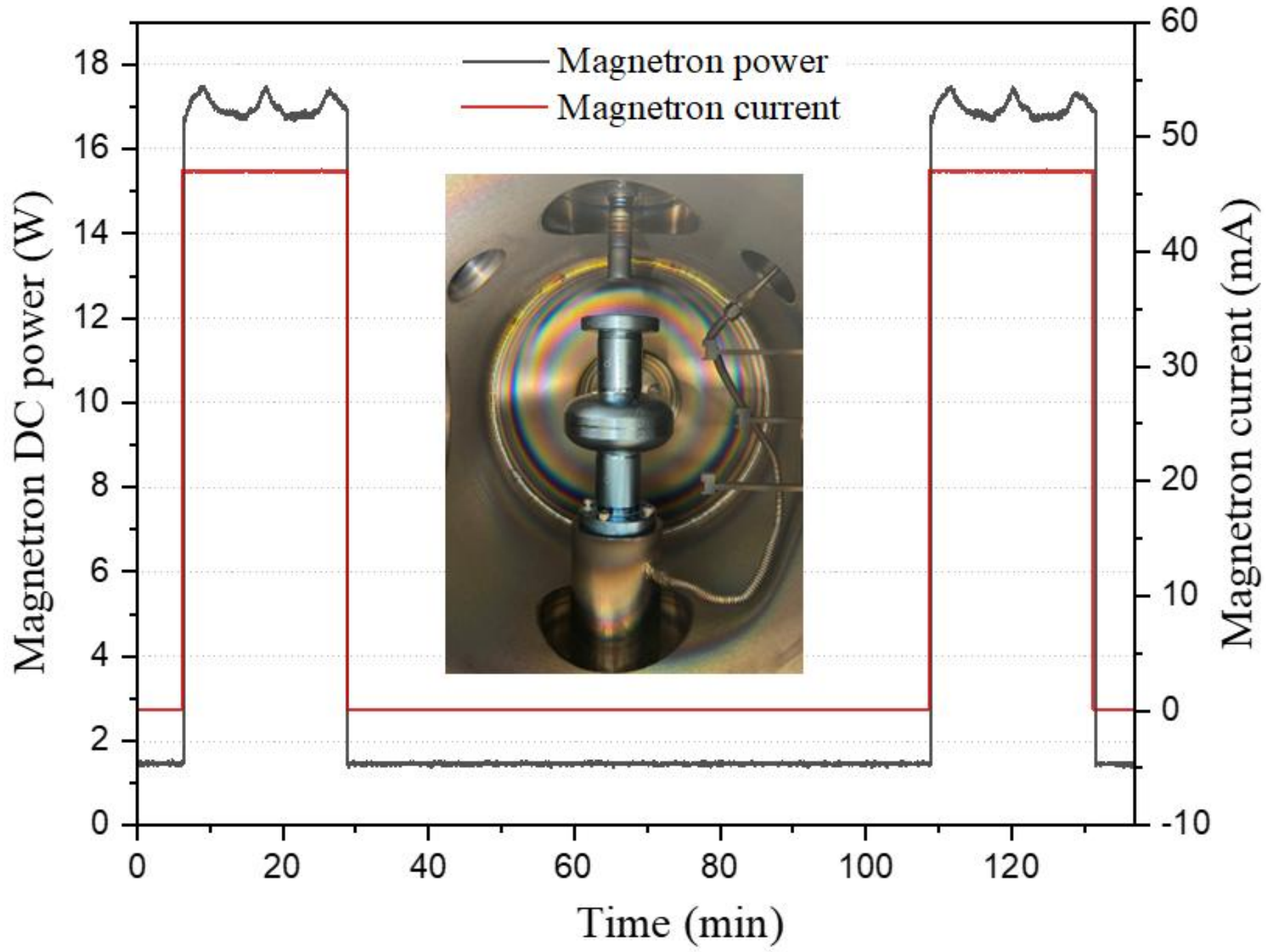

**Fig. 12.** Magnetron power (DC) during the Nb-Sn coating into 2.6 GHz Nb SRF cavity via co-sputtering deposition process. A fixed magnetron discharge current of 47 mA was used during the deposition, including an 80-min in which the discharge is turned off to allow cooling of the target after every two passes of the magnetron across the cavity. The inset shows an anodized 2.6 GHz Nb SRF cavity positioned in the cylindrical magnetron sputtering chamber for Nb-Sn coating via the co-sputtering. The corrugated stainless steel tube supplies Ar gas into the cavity base for sputter deposition.

Upon completion of co-sputtering, the coated cavity was removed from the cylindrical magnetron sputtering system, transferred to an ISO4 cleanroom at Jefferson Lab, and subjected to a high pressure rinsing at 50 bar water pressure. The Nb-Sn coated cavity was then transferred to Fermilab for annealing in a high vacuum furnace. The furnace consists of a 12″ × 12″ × 18″ hot zone enclosed by molybdenum heat shields. The cavity was placed on niobium supports inside a niobium box. The assembly was placed in the hot zone, and the furnace was evacuated to a base pressure of $8.8 \times 10^{-7}$ Torr. Before the annealing process, the furnace was degassed at about 200 °C for 60 h, and the furnace pressure reached $4.9 \times 10^{-7}$ Torr. The annealing process included a temperature ramp at about 3 °C/min to 600 °C and holding it for 6 h. The temperature was then ramped at about 12 °C/min to 950 °C, followed by annealing at 950 °C for 1 h, as presented in Fig. 13. Figure 14(a) shows the as-deposited coated cavity surface, which appeared visually uniform and consistent across the cavity. No visible signs of film peeling or particle adhesion were observed on the surface. After annealing, the cavity surface displayed a uniform, slightly whitish coloration across its entire surface, as shown in Fig. 14(b).

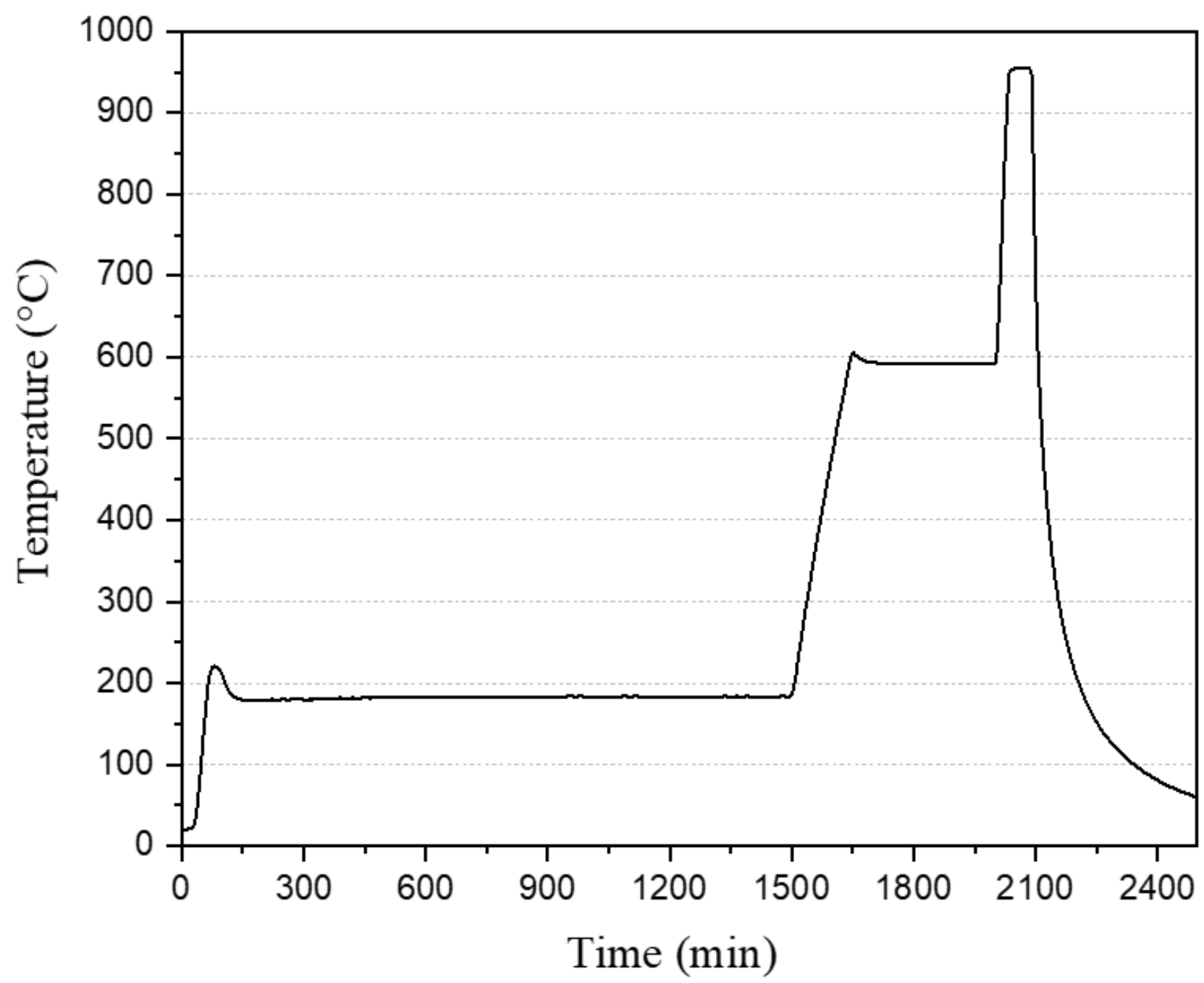

**Fig. 13.** Temperature profile of the two-step annealing process for the Nb-Sn coated cavity via co-sputtering, consisting of an initial annealing at 600 °C for 6 h, followed by a second annealing at 950 °C for 1 h.

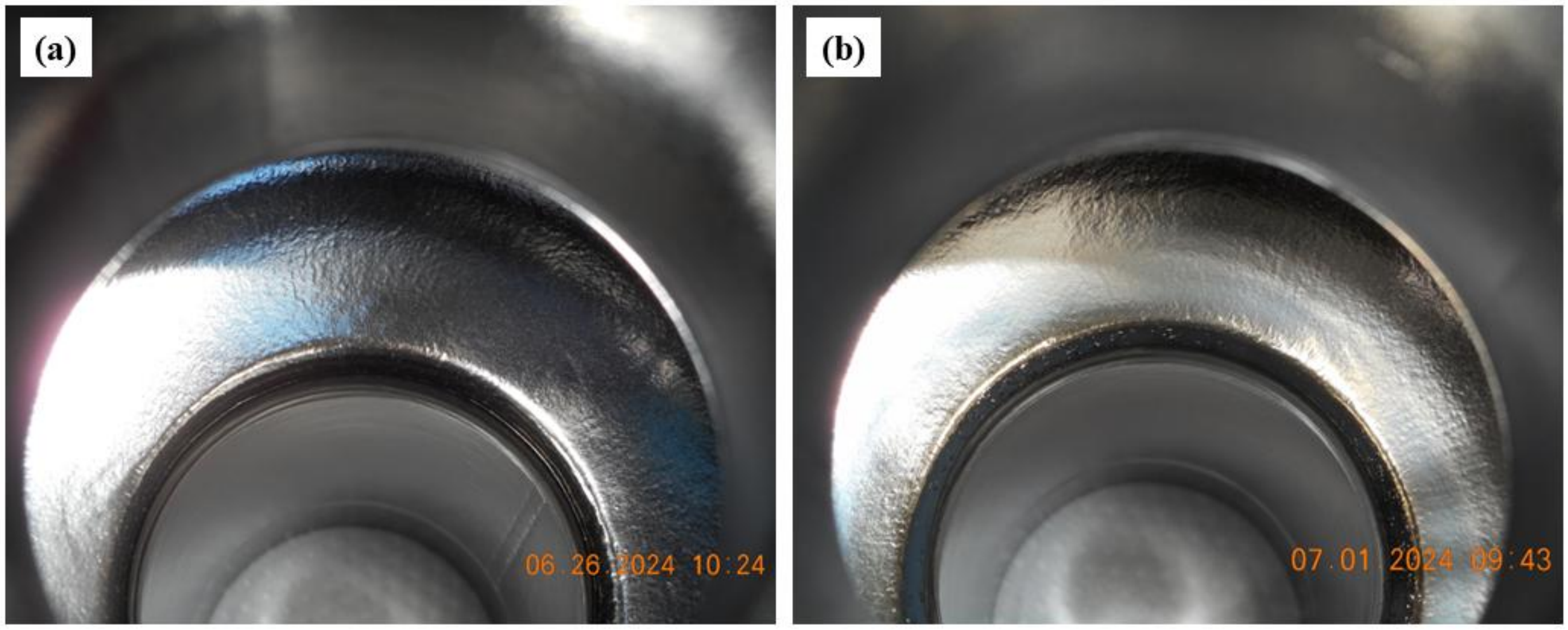

**Fig. 14**. Photograph of the coated cavity surface: (a) As-deposited and (b) after annealing at 600 °C for 6 h, followed by annealing at 950 °C for 1 h.

### 3.4.1. RF performance of $Nb_3Sn$ coated cavity

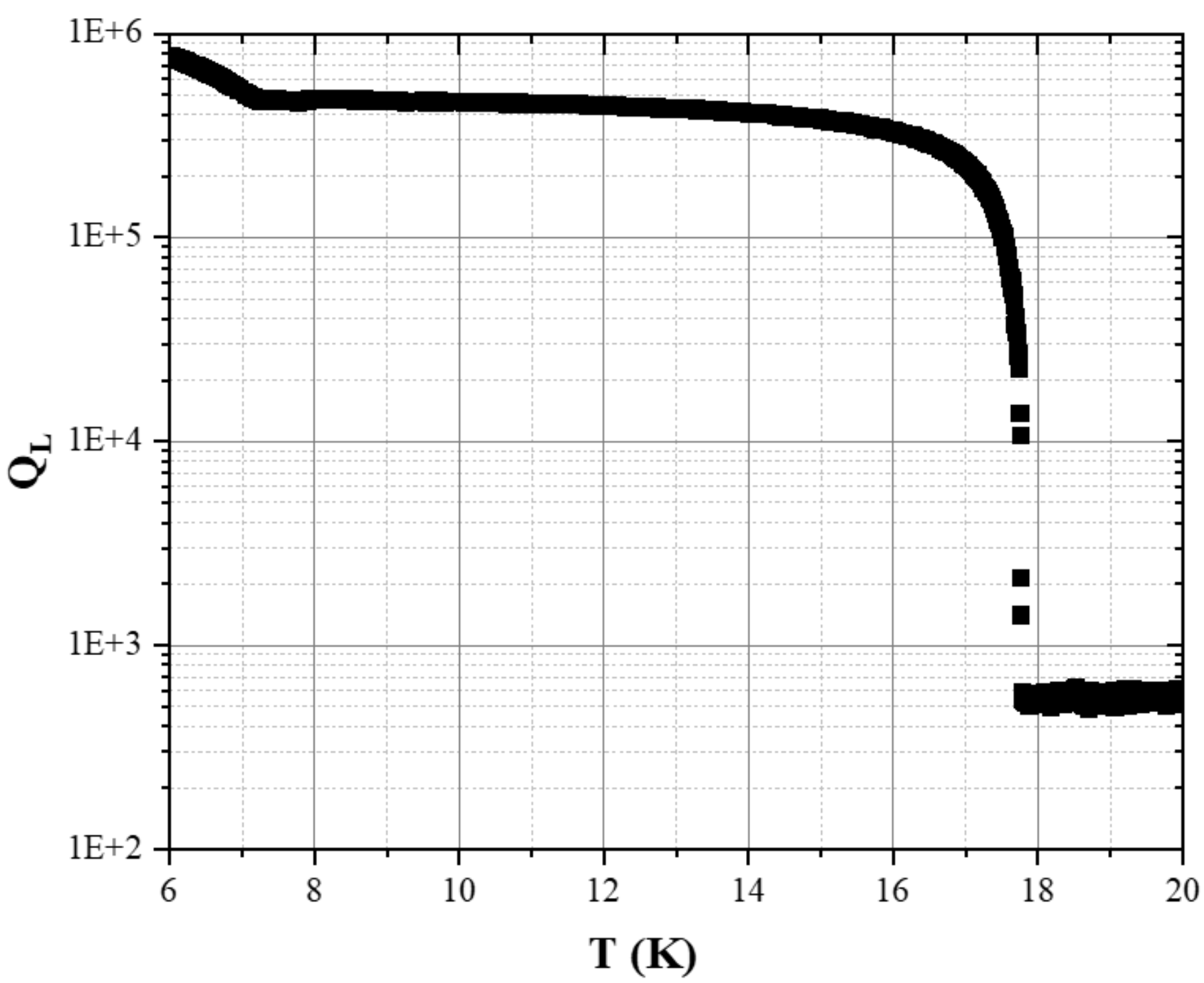

**Fig. 15.** $Q_L$ versus T data showing the superconducting transition at about 17.8 K due to $Nb_3Sn$ coating. A second transition observed at about 7.4 K is attributed to Sn deficient areas in $Nb_3Sn$ films at the equator region.

The annealed cavity was removed from the furnace and prepared in a cleanroom for cryogenic RF testing. The cavity underwent high pressure rinsing, drying, and was then assembled in the ISO4 cleanroom for the vertical test stand (VTS) system at Fermilab [40,41]. The cavity was slowly cooled to cryogenic temperatures to avoid magnetic flux trapping caused by thermal currents generated by non-uniform temperature across the cavity during the superconducting transition. The temperature gradient across the cavity was about 0.1 K when the average temperature was about 18 K. The loaded quality factor ($Q_L$) of the cavity was measured with a network analyzer and found to be $1.3 \times 10^{-6}$ at 4.4 K.

A network analyzer was then used to measure the resonant frequency and loaded $Q_L$ as a function of temperature during dewar warm-up. $Q_L$ as a function of temperature is shown in Fig.15. The temperature dependence of $Q_L$ exhibits two transitions: one at about 7.4 K and the other at about 17.8 K. The transition at 17.8 K is attributed to the superconducting transition temperature of the $Nb_3Sn$ layer. The second superconducting transition at about 7.4 K is attributed to potential Sn-deficient $Nb_3Sn$ in the equator region of the annealed cavity.

### 3.4.2. Light Sn recoating onto $Nb_3Sn$ coated cavity and RF performance

After removing the cavity from the cryogenic dewar, it underwent the standard preparation steps for vacuum furnace annealing. The light recoat process for the cavity followed the procedure described in [35]. The vacuum inside the retort, where the cavity resides during coating, was $2 \times 10^{-7}$ Torr at the end of the degassing step and about $1 \times 10^{-6}$ Torr during the 1000 °C coating step. Following the coating process, the cavity was removed from the furnace and prepared for VTS testing, as described earlier.

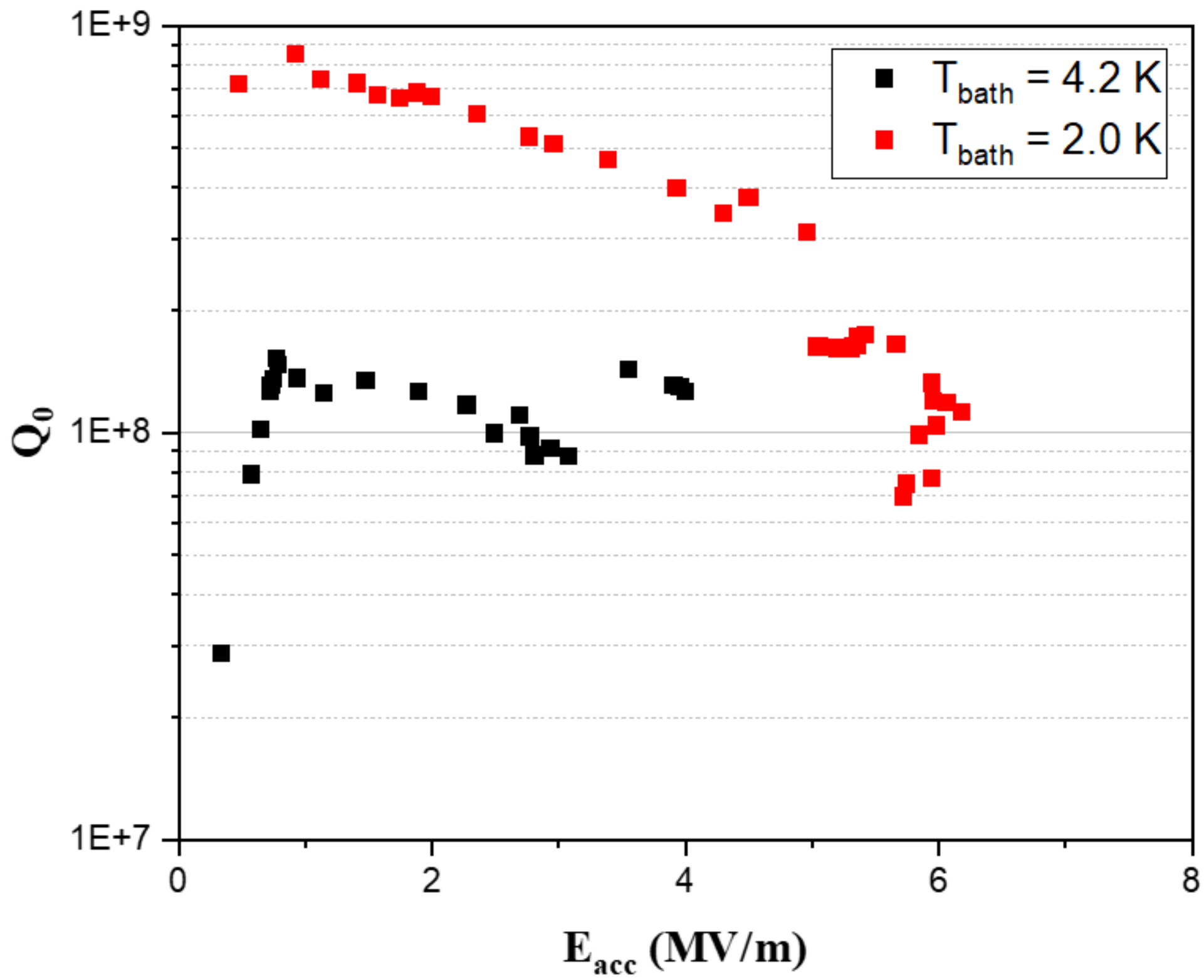

Fig. 16. $Q_0$ versus $E_{acc}$ at 4.2 K and 2.0 K following the light Sn recoating of 2.6 GHz Nb cavity. Before Sn recoating, about 1.5 µm $Nb_3Sn$ was deposited inside the cavity by co-sputtering, followed by annealing at 600 °C for 6 h and then at 950 °C for 1 h.

Figure 16 presents the $Q_0$ versus $E_{acc}$ performance of the Sn recoated 2.6 GHz Nb cavity during the cryogenic RF test. At $T_{bath}$ = 4.2 K, $Q_0$ exhibited a two-order-of-magnitude increase compared to the pre-recoating state. Under these conditions, $Q_0$ reached a maximum value of $1.5 \times 10^8$ at low fields, with $E_{acc}$ limited to 4 MV/m. At $T_{bath}$ = 2.0 K, $Q_0$ was $8.5 \times 10^8$ at low fields, while $E_{acc}$ was constrained to 6.1 MV/m. $T_c$ of the Sn recoated cavity was measured to be 17.6 K, attributed to the $Nb_3Sn$ layer.

The enhanced cryogenic RF performance of the cavity is attributed to the improved microstructural uniformity and improved composition of the $Nb_3Sn$ coating following the light Sn recoating treatment. Initially, the annealed $Nb_3Sn$ coating exhibited surface voids on the equator region, see Figure 6, which likely led to increased surface resistance ($R_s$), resulting in a very low $Q_0$, which precluded the measurement of $Q_0$ as a function of the accelerating gradient. The light Sn recoating may have eliminated these voids, similar to the case of flat Nb samples as seen in Fig 10, thereby reducing the residual surface resistance ($R_{res}$), resulting in enhanced $Q_0$. Note that the treatment has shown the healing of microstructural cracks, reported in [35]. Sn vapor during the treatment is also expected to improve the $Nb_3Sn$ composition, which had suffered Sn loss during annealing. Despite these improvements, the performance of the co-sputtered cavity remains inferior to that coated with $Nb_3Sn$ using the conventional Sn vapor diffusion method, which has demonstrated $Q_0$ values as high as $2 \times 10^{10}$ at 2.0 K [42]. Several factors may have contributed to the lower performance observed in the co-sputtered $Nb_3Sn$ cavity. One potential cause is microstructural void formation in the equator area during the co-sputtering, leading to Sn-loss during post-sputtering annealing, which could result in poor stoichiometry and thinner coating, negatively affecting RF performance. Due to limitations in the sputtering geometry, the coating uniformity can also vary in the cavity cell because of its curvature. These limitations may be eliminated by improving the process design in the future, for example, by providing an appropriate voltage bias or a heating set-up for the cavity during sputtering.

## 4. Conclusion

A co-sputtering deposition technique has been developed and optimized for $Nb_3Sn$ coating on a 2.6 GHz Nb SRF cavity using a cylindrical magnetron sputtering system. The optimal target arrangement and deposition conditions were determined using a mockup cavity replicating a 2.6 GHz Nb SRF cavity geometry. Co-sputtered Nb-Sn films with an atomic Sn content of 32 – 42% in the as-deposited samples, representing the beam tube and equator positions of the cavity, formed $Nb_3Sn$ with good crystallinity after annealing at 950 °C for 3 h. Multiple annealing approaches were explored to improve the microstructure and compositional uniformity of the annealed films. A 1.5 μm thick Nb-Sn film was deposited onto the inner surface of a 2.6 GHz Nb SRF cavity, and the coated cavity was annealed using a two-step annealing process: first at 600 °C for 6 h, followed by 950 °C for 1 h. Cryogenic RF testing of the $Nb_3Sn$-coated cavity demonstrated a superconducting transition temperature of 17.8 K, indicating the formation of the $Nb_3Sn$ coating on the cavity surface. Finally, a light Sn recoating was applied to the $Nb_3Sn$ coated cavity to improve the coating quality, which resulted in a significant enhancement in RF performance, achieving a low-field $Q_0$ up to $8.5 \times 10^8$ at 2.0 K.

### Acknowledgment

This manuscript is based upon work supported by the U.S. Department of Energy, Office of Science, Office of Accelerator R&D and Production, under contract No. DE-SC0022284 and the U.S. Department of Energy, Office of Science, Office of Nuclear Physics under contract DE-AC05-06OR23177 with Jefferson Science Associates, including supplemental funding via the DOE Early Career Award to G. Eremeev. This manuscript has been authored by FermiForward Discovery Group, LLC under Contract No. 89243024CSC000002 with the U.S. Department of Energy, Office of Science, Office of High Energy

Physics. The authors acknowledge Olga Trofimova from the Applied Research Center Core Labs at the College of William & Mary for her help with FESEM imaging.